\documentclass{article}

\usepackage{microtype}
\usepackage{graphicx}
\usepackage{subfigure}
\usepackage{booktabs}
\usepackage{hyperref}
\usepackage{longtable}
\usepackage{array}
\usepackage{ragged2e}
\usepackage{pdflscape}
\usepackage{booktabs}
\usepackage{colortbl}
\usepackage{xcolor}

\usepackage{xurl}

\usepackage[accepted]{icml2026}

\usepackage{amsmath}
\usepackage{amssymb}
\usepackage{mathtools}
\usepackage{amsthm}

\usepackage[capitalize,noabbrev]{cleveref}

\theoremstyle{plain}

\theoremstyle{definition}

\theoremstyle{remark}

\icmltitlerunning{Principles and Guidelines for RCTs in AI Evaluation}

\begin{document}

\twocolumn[
\icmltitle{Principles and Guidelines for Randomized Controlled Trials in AI Evaluation}

\begin{icmlauthorlist}

\icmlauthor{Christopher Kelly\textsuperscript{\textdagger\textdaggerdbl}}{independent}
\icmlauthor{Angelica Chowdhury\textsuperscript{\textdaggerdbl}}{independent}
\icmlauthor{Alexandra Campili\textsuperscript{\textdaggerdbl}}{independent}
\icmlauthor{Bimpe Ayoola\textsuperscript{\textsection}}{bimpe}
\icmlauthor{Devin Barbour\textsuperscript{\textsection}}{devin}
\icmlauthor{Thomas Chen Dawson\textsuperscript{\textsection}}{thomas}
\icmlauthor{Ze Shen Chin\textsuperscript{$\circ$}}{comp_om,comp_aisl}
\icmlauthor{Rokas Gipiškis\textsuperscript{$\circ$}}{comp_aisl,rokas}

\end{icmlauthorlist}

\icmlaffiliation{comp_aisl}{AI Standards Lab}
\icmlaffiliation{comp_om}{Oxford Martin AI Governance Initiative}
\icmlaffiliation{rokas}{Vilnius University}
\icmlaffiliation{bimpe}{Dalhousie University}
\icmlaffiliation{devin}{Northeastern University London}
\icmlaffiliation{thomas}{University of Pennsylvania}
\icmlaffiliation{independent}{Independent}

\icmlcorrespondingauthor{Rokas Gipiškis}{rokas@aistandardslab.org}

\icmlkeywords{AI Safety, Evaluation, RCT, Uplift Studies, GPAI}

\vskip 0.3in
]

\printAffiliationsAndNotice{\textsuperscript{\textdagger}Primary Author, writing the original draft and core contribution; \textsuperscript{\textdaggerdbl}Guidelines Contributor, contributing to guideline development; \textsuperscript{\textsection}Project Participant, providing feedback and participating in SPAR (Supervised Program for Alignment Research) meetings; \textsuperscript{$\circ$}Project Mentor.}

\begin{abstract}
This work establishes a framework for standardizing AI evaluation RCTs (sometimes called human uplift studies). Drawing on established practices from disciplines with established RCT traditions, including software engineering, economics, clinical and health sciences, and psychology, we synthesize five principles drawn from established validity frameworks and open-science standards on transparency, repeatability, and verification, which together serve as the conceptual foundation for 33 actionable guidelines adapted for AI evaluation RCT contexts, expressed as requirements with rationales, implementation instructions, and evidence bases. We position the principles and guidelines as serving three key roles for AI evaluation RCTs: a design tool for planning studies, an evaluation rubric for assessing existing work, and a blueprint for standard setting as the field converges on norms. AI evaluation research currently lacks common standards and shared vocabulary for producing cumulative, comparable, policy-ready evidence. This framework is a contribution toward that foundation, providing evaluative criteria and a shared conceptual language alongside actionable guidelines.

\end{abstract}

\section{Introduction}

The credibility of empirical research has come under sustained scrutiny over the past decade. Large-scale replication efforts in psychology found that fewer than half of published findings could be reproduced \cite{nosek2015}. Systematic reviews in medicine have documented pervasive heterogeneity in trial designs, small sample sizes, lack of randomization, and inadequate transparency \cite{zhao2021}. Software engineering has grappled with similar challenges, prompting the development of community standards to codify methodological requirements for experimental studies \cite{sigsoft2020}. These problems are not limited to any single field. They reflect structural incentives and institutional gaps that undermine the production of reliable knowledge across the sciences.

Against this backdrop, the evaluation of advanced AI systems\footnote{Throughout this paper we use \textit{AI} to refer to advanced general-purpose AI systems, as these motivate our examples and framing. The literature uses several terms for this category, including \textit{general-purpose AI} (GPAI) \cite{rand2025,aisi2024}, \textit{frontier AI} \cite{anderljung2023,wijk2024,buhl2025}, \textit{foundation models} \cite{bommasani2021}, and \textit{large language models} (LLMs) \cite{zhao2023}. However, the principles and guidelines themselves are methodological rather than system-specific: they apply to RCTs evaluating any AI 
system, whether general-purpose or domain-specific, and whether the object of evaluation is a model or a broader system \cite{sun2026}; our framework applies across these usages.} has emerged as a domain with virtually no established standards of best practice \cite{aisi2024}. Benchmark evaluations, leaderboards, and audits do not amount to a robust evaluation ecosystem, and they fail to provide assurances regarding AI system performance across different domains or for different user groups \cite{weidinger2025}. The machine learning field has historically overlooked the importance of evaluation practices, and the rate at which general-purpose AI capabilities are progressing compounds this problem by increasing the risk of unbalanced assessments \cite{weidinger2025}. Overestimating AI capabilities can lead to frayed trust, hasty deployment, and over-reliance on systems whose performance has not been sufficiently verified. Underestimating them can result in complacency, underinvestment in safety, and missed opportunities for beneficial applications.

The problem is further complicated by the use of inconsistent terminology. In the scholarly domain, the concept of ``uplift'' is formally defined and operationalized differently across disciplines. Sometimes it is confused with ``treatment effect'' \cite{uplift2024}, ``incremental impact'' \cite{gutierrez2017}, ``performance evaluation'' \cite{verhelst2023}, and ``gain'' \cite{jaroszewicz2017}. Although these terms are frequently used interchangeably, they do not mean exactly the same thing and have important differences. A treatment effect can indicate negative, null, or positive change, whereas uplift implies improvement and therefore carries a positive directional meaning. 
``Program Evaluation'' often refers to a much less rigorous version of evaluation than an RCT or quasi-experiment and does not reference treatment effects, and it is more akin to a simple before and after. Among leading AI organizations, definitions contradict one another. The RAND GPAI paper employs a broad definition encompassing any increases in human performance from AI access \cite{rand2025}, while the UK AI Safety Institute frames uplift studies as assessing how AI systems might be used by bad actors to carry out real-world harmful tasks \cite{aisi2024}. Anthropic's usage parallels this focus on CBRN risks but lacks a methodological definition \cite{anthropic2025}. OpenAI rarely employs the term and uses it differently when it does, describing changes in model performance across model versions rather than changes in human performance \cite{openai2024}. This terminological inconsistency undermines the interdisciplinary research and policy conversations that are essential for AI safety and governance.

The consequences extend beyond academic confusion. Without clear standards, we risk inconsistent methodologies and unverified claims that limit the validity of conclusions and make it difficult to reproduce results \cite{mishra2020}. Policymakers and governance bodies cannot aggregate findings across studies when those studies use different definitions, different outcome measures, and different analytical approaches. Technical governance decisions about deployment thresholds, capability assessments, and safety requirements depend on evidence that cannot be trusted if the underlying studies are underpowered, poorly designed, or inadequately documented. The CSET explainer on AI safety evaluations acknowledges the inherent difficulty in performing contextual safety evaluations due to hard-to-standardize human variables, such as tester expertise, which can complicate establishing clear cause-and-effect relationships and comparing results across different studies \cite{friedland2025}.

 In this work, we define an \emph{AI evaluation RCT} as a randomized controlled trial measuring the causal effect of AI access or use on a given outcome. Unlike benchmark evaluations, which assess model outputs against fixed test sets, AI evaluation RCTs randomize human participants to AI-access conditions and measure causal effects on human behavior, performance, or wellbeing. We focus specifically on human performance outcomes because many capability thresholds currently use ambiguous language about increasing human capabilities or automating expert-level work without specifying whether the human baseline is AI-assisted \cite{rand2025}. Our framework is domain-agnostic and applies across task domains and settings, though our focus is specifically on experimental designs that enable causal inference.

This work adopts the Shadish-Cook-Campbell four-validity framework \cite{shadish2002}, extends it with a fifth principle on transparency and repeatability adapted from the TOP Guidelines \cite{top2025}, and operationalizes all five into 33 guidelines adapted for AI evaluation RCT contexts, drawing on established standards across multiple disciplines. This paper intends to synthesize established best practices from fields with mature RCT traditions, generate novel AI-specific extensions where existing frameworks fall short, and package the result as an actionable resource for a research community that currently lacks such a resource. Components not present in prior frameworks include an AI-specific catalog of validity threats, an inferential scope table describing what each validity principle does and does not license a researcher to conclude, and a structured critique of inconsistent uplift terminology alongside a glossary standardizing key terms across disciplines. The framework is also designed for interdisciplinary use, providing shared vocabulary and validity criteria that researchers, practitioners, and engineers from different backgrounds can apply consistently, making findings comparable across studies.

\section{Related Work and Gaps}

Several efforts have attempted to establish standards for AI evaluation, though none adequately address the unique challenges of AI evaluation RCTs.

STARE-HI \cite{talmon2009} represents an early effort to structure reporting standards for evaluation studies of complex digital systems in healthcare. It emphasizes that evaluations should capture not just technical performance, but also context, user interaction, workflow integration, and unintended consequences. However, it predates the rise of machine learning and adaptive AI systems and therefore lacks provisions for model evolution, data drift, prompting, and domain-specific constraints within healthcare.

CONSORT (Consolidated Standards of Reporting Trials) \cite{schulz2010} is a widely adopted checklist for reporting randomized controlled trials, designed to ensure transparent and complete reporting of trial design, conduct, and outcomes. CONSORT-AI \cite{liu2020} extends this framework for clinical AI interventions, adding 14 new items to address algorithm versioning, training data provenance, model update policies, and human-AI interaction during clinical trial execution. SPIRIT (Standard Protocol Items: Recommendations for Interventional Trials) \cite{chan2013} provides complementary guidance for clinical trial protocol development, specifying what information should be included before a trial begins. SPIRIT-AI \cite{rivera2020} adapts this for AI contexts, requiring pre-specification of the AI system's intended use, input data, human-AI interface, and plans for handling model updates. Both frameworks assume medical contexts with static interventions, clinical outcomes, and well-established regulatory oversight. 

STREAM \cite{mccaslin2025} introduces a structured framework for reporting evaluations of AI models, initially focusing on chemical and biological threat assessment. It defines criteria across six domains aimed at improving reproducibility, comparability, and accountability. However, STREAM explicitly excludes AI evaluation RCTs, focusing instead on benchmark evaluations.

The RAND Perspective \cite{rand2025} is a set of preliminary best practice recommendations that identifies challenges in broader AI evaluation efforts: weak construct validity, inconsistent measurement of real-world risk, and insufficient reproducibility. RAND proposes cross-cutting best practices drawn from disciplines such as psychology, economics, and biology, organized around the four life-cycle stages of evaluation: design, implementation, execution, and documentation. Although preliminary, the RAND framework directly calls for clearly set-out best practices guiding evaluation of GPAI systems.

Complementing these normative frameworks, \citet{paskov2026} provide empirical grounding through interviews with 16 expert practitioners who have conducted human uplift RCTs across domains including biosecurity, cybersecurity, education, and labor. Their findings document a recurring tension between standard causal inference assumptions and the distinctive properties of frontier AI systems: rapidly evolving models, shifting user baselines, heterogeneous AI literacy, and porous real-world settings that strain internal, external, and construct validity. Where Paskov et al.\ identify and map practitioner-experienced challenges, our work provides a normative framework of principles and operationalized guidelines designed to address precisely those challenges.

Several gaps remain when adapting these frameworks to AI evaluation RCT contexts. Treatment definition and versioning is inadequately addressed because many existing guidelines assume static interventions, while AI systems evolve rapidly through updates, prompt changes, or newly fine-tuned variants. Spillover and interference are similarly neglected, as existing frameworks do not account for the potential for unsanctioned model use in multi-user AI environments, which can violate classical RCT assumptions. Human-AI interaction reporting is largely absent from existing frameworks, which do not provide guidance on documenting user instructions, degree of autonomy, or how outputs were modified or accepted. Construct validity is underspecified for AI contexts, where standard performance metrics, such as accuracy or completion time, may poorly reflect intended outcomes like reasoning quality, trust, or creativity \cite{salaudeen2025}. Finally, reproducibility is under-addressed: many AI models are closed-source, meaning existing transparency provisions designed around model weights and training data do not translate to API-based systems, where reproducibility depends on logs, prompts, and metadata \cite{rand2025}.

\section{Principles \& Guidelines Overview}

Our work rests on five principles: four drawn from the Shadish-Cook-Campbell validity framework \cite{shadish2002} and a fifth adapted from the Transparency and Openness Promotion (TOP) Guidelines \cite{top2025}. These principles function as interpretive tools for determining what claims a given study can and cannot support. \hyperref[tab:inferential]{Appendix~B} provides a walk through of what each principle enables you to conclude, does not enable you to conclude, and contains examples of appropriate language for discussing that principle in a paper. Each principle also addresses a distinct dimension of what can undermine the credibility and validity of research conclusions. \hyperref[tab:threats]{Appendix~C} provides examples of threats to each of these principles, drawn from Shadish, Cook, and Campbell \cite{shadish2002} and extended with AI evaluation RCT-specific examples. It includes the name of the threat, a description of the threat, and an example of what that could look like in AI evaluation RCTs. This framework thus helps readers calibrate confidence in reported findings and identify the specific inferential boundaries that a study's methods impose.

\subsection{Construct Validity}

The construct validity principle asks whether outcome measures in an AI evaluation RCT actually capture the underlying constructs of interest, such as coding competence, diagnostic reasoning, or writing quality, rather than narrow proxies that are easy to score but misaligned with the real capability question. This follows the classic view of construct validity as the degree to which a measure supports appropriate inferences about an underlying attribute \cite{cronbach1955, shadish2002}.

In settings involving AI, construct validity is strained by both the fact that AI tools can change the nature of the task itself and the temptation to focus on metrics like speed or token counts that models can optimize for without genuinely enhancing human skill. We highlight threats such as construct underrepresentation, where outcomes only cover the slice of performance that the model makes easier, and construct-irrelevant variance, where AI boosts scores through superficial fluency or verbosity. When construct validity is weak, a study may show that scores increased with AI, but it cannot credibly claim that human capability in the targeted domain has improved.

\subsection{Internal Validity}

The internal validity principle concerns whether observed differences between conditions can be credibly attributed to access to a specific AI system rather than to confounders \cite{shadish2002}. In AI evaluation RCTs, internal validity is threatened not only by familiar issues such as selection, attrition, and concurrent external events, but also by AI-specific phenomena: contamination of control groups through unsanctioned model use, spillovers and interference within teams or classes, model updates during the study, and differential compliance with AI usage instructions.

Our treatment of this principle catalogs these threats and emphasizes design choices, such as aligning the unit of randomization with the interference structure, monitoring compliance, and pinning model versions. 

Internal validity is particularly consequential in AI evaluation research because the evidence from these studies feeds directly into deployment decisions, impact assessments, safety thresholds, public policy, and governance frameworks, all of which require causal evidence rather than observed correlations. Knowing that AI use and performance improvement co-occur in an organization does not necessarily establish that AI caused the improvement. A poorly designed study with selection effects, concurrent organizational changes, or differential motivation among AI adopters can produce the same pattern without any causal relationship. An RCT design addresses this by construction, but only if the randomization is appropriately carried out. The AI-specific threats named above can each compromise causal identification even in a nominally randomized design. When internal validity fails, a study may still describe what happened under specific conditions, but it cannot support claims about what AI access caused.

\subsection{External Validity}

The external validity principle focuses on to whom, for what tasks, and in which settings the findings of an AI evaluation RCT are intended to generalize. External validity is often described as the question of whether a causal relationship holds over variations in persons, settings, treatments, and outcomes \citep{shadish2002}.

In our setting, this includes careful definition of the target population (novice versus expert practitioners, AI-naive versus AI-literate users), the task and domain, and the organizational and technical context in which AI is used. Threats include heavy reliance on convenience or WEIRD (Western, Educated, Industrialized, Rich, and Democratic) samples, short artificial tasks that do not reflect real work, deployment contexts that differ radically from study conditions, and AI systems that are loosely specified or rapidly changing.

External validity is especially pressing in AI evaluation research because the systems under study change rapidly, the populations using them are heterogeneous in ways that matter substantially for outcomes, and the gap between study conditions and real deployment contexts is often large. A study conducted with a specific model version, on a narrow task, with a relatively AI-literate participant pool, may produce findings that do not transfer to a different model, a different task domain, or users with less experience with AI tools. However, the temptation to generalize broadly is strong given the policy and deployment stakes. Each study should therefore specify its inferential scope explicitly, including for which population, task, organizational context, and AI implementation the findings are intended to hold, and what additional evidence would be needed to extend those claims. Findings from single studies should be treated as informative about a specific configuration, not as evidence about AI effects in general.

\subsection{Statistical Conclusion Validity}

The statistical conclusion validity principle addresses whether quantitative evidence is sufficient and appropriately interpreted for any claimed effect. Shadish, Cook, and Campbell treat statistical conclusion validity as distinct from internal validity, focusing on issues such as low power, unreliability, and violations of statistical assumptions \cite{shadish2002}.

In AI evaluation RCTs, this encompasses evidentiary thresholds for novel claims, power and precision (especially for heterogeneity and interaction effects), appropriate treatment of multiple outcomes and analyses, and a shift from dichotomous significance testing toward estimation and graded evidence. We set $\alpha = 0.005$ for novel causal claims, emphasize effect size estimation with confidence intervals, and require sensitivity analyses \citep{benjamin2018}. Threats include underpowered designs, noisy or low-reliability measures, extensive researcher degrees of freedom without pre-specification, and over-interpretation of fragile or marginal results. When statistical conclusion validity is weak, a study’s statistical evidence may not support its conclusions: it may falsely indicate an effect, fail to detect a real one, or produce an effect-size estimate too imprecise or unstable to trust. This can remain true even when other aspects of the study, such as randomization, implementation, or construct validity, are relatively strong.

\subsection{Transparency, Repeatability, and Verification}

The transparency, repeatability, and verification\footnote{The term ``reproducibility'' has historically been used as a synonym for several distinct concepts---including reproducibility, robustness, replicability, repeatability, credibility, and trustworthiness---leading to ambiguity and miscommunication across fields. We follow \citet{nosek2026glossary} in using \textit{repeatability} as the umbrella term for the constellation of concepts that assess whether research findings are unchanged when different steps of the research process are repeated. Within repeatability, we distinguish: \textit{reproducibility} (same data, same analysis), \textit{robustness} (same data, different analysis), and \textit{replicability} (new data, same question). These three dimensions are empirically partially independent---knowing a finding is reproducible provides limited information about whether it is replicable \citep{nosek2026glossary,aczel2026}---and should be treated as distinct forms of evidence rather than rungs on a single ladder. Readers accustomed to using ``reproducibility'' as a general term are encouraged to consult \citet{nosek2026glossary} for a full discussion of why these distinctions matter for scientific communication.} principle asks whether other researchers, practitioners, and governance bodies can see what was done, repeat the analyses, and independently assess the credibility of findings. This principle aligns with the Transparency and Openness Promotion guidelines and broader work on scientific repeatability \citep{nosek2015,top2025,nosek2026pnas}.

GPAI uplift studies are particularly vulnerable to opacity. Key details about model family and version, prompts and settings, fine-tuning data, interface design, usage policies, compliance monitoring, and deviations from original research and analysis plans are often under-reported or omitted entirely. Transparency and repeatability are especially critical in this context for four reasons. First, AI evaluation RCT research informs high-stakes decisions about deployment, workforce impacts, and policy, and must establish credibility through verifiable, repeatable findings. Evidence from the social and behavioral sciences suggests these are not minor concerns: only about half of published claims replicate with the same pattern when tested with new data, and only about one-third of reanalyses closely match original results \citep{tyner2026,aczel2026}. Second, the many hidden degrees of freedom in AI evaluation RCTs---model specifications, prompt engineering, preprocessing decisions, analytical choices---create substantial opacity without active transparency efforts. Third, rapid AI system evolution means that without detailed documentation, studies become impossible to interpret or reproduce as systems change. Fourth, research errors are common: approximately one quarter of articles in fields with data and code sharing requirements contain non-trivial coding errors \citep{brodeur2026}, and transparency enables their detection and correction. Critically, journals requiring data sharing empirically produce more reproducible results \citep{miske2026}, suggesting that the specific practices mandated by this framework are not merely normative recommendations but empirically supported interventions.

We implement a four-level transparency, repeatability, and verification framework in which the strength of causal claims should scale with verifiability. The first three levels form a genuine hierarchy in which each builds on the previous: \textbf{Level 1, Disclosure}, requires that researchers report whether each open science practice was conducted and, if so, how. This enables critical readers to assess what was and was not done. \textbf{Level 2, Sharing}, requires that the outputs of each practice are deposited in accessible locations and linked in the paper, so that independent reuse and reproduction become practically possible; data sharing requirements empirically improve reproducibility \citep{miske2026}. \textbf{Level 3, Verification}, requires that an independent party (e.g. a journal, auditor, or registered report process) confirms that what was claimed at Levels 1 and 2 is genuinely and completely fulfilled, including that code runs and produces the reported outputs and that pre-registrations are complete and timestamped before data collection. \textbf{Level 4, Repeatability}, groups three partially independent dimensions that together provide the strongest transparency, repeatability, and verification evidence \citep{nosek2026glossary,aczel2026}: \textit{Reproducibility} (same data, same analysis) requires that independent parties rerun the analysis on shared data and code and confirm reported results as a minimum form of repeatability, but insufficient alone. \textit{Robustness} (same data, different analysis) requires that independent parties apply alternative justifiable analytical choices and assess whether qualitative conclusions hold; large-scale evidence shows that while 74\% of reanalyses reach the same qualitative conclusion as originals, only 34\% closely match the original quantitative estimates \citep{aczel2026}. \textit{Replicability} (new data, same question) requires that independent teams test the same research question with new participants and settings; replication effect sizes in the social and behavioral sciences are typically less than half the magnitude of original estimates \citep{tyner2026}, suggesting that single-study AI evaluation RCT findings warrant corresponding caution. These three dimensions are empirically near-zero correlated and should not be assumed to imply one another \citep{nosek2026glossary}.

Studies meeting only Level 1 warrant cautious interpretation; studies achieving Level 3 verification and Level 4 repeatability assessment provide substantially stronger grounds for belief in reported effects. The appropriate strength of causal claims should scale with which dimensions of repeatability have been assessed.

\subsection{Relationships Among the Principles}

The five principles are parallel in design, each addressing a distinct dimension of what can undermine the credibility of a study's conclusions. No single principle is sufficient on its own. Rather, the credibility of a study's findings depends on how well all five are addressed together. For different studies, different types of validity may be more or less relevant. And there are more types of validity than the five discussed in this paper. AI research has often focused on benchmarks (i.e.\ task-oriented evaluation or tests), rather than types of validity. Recently, AI research has begun to focus more on validity related to benchmarks, especially construct validity \citep{salaudeen2025, wang2026}. While this focus on construct validity is important, it may have narrowly focused the field in a way that neglects the importance of other types of validity. This is particularly relevant since other types of validity are important for RCTs and quasi-experiments. Other fields have encountered similar phenomena. Vazire et al.\ argue that psychology's credibility movement over-indexed on statistical conclusion validity, via the replication crisis focus, at the expense of the other validity types, and that improving research quality requires attention to all of them \citep{vazire2022}.

Despite being parallel, the principles interact in ways that matter for how findings should be interpreted. Gaps in different principles produce distinct types of problems that cannot substitute for or cancel out one another. For example, a causally identified effect on a poorly chosen outcome measure is a credible answer to the wrong question. Without adequate transparency and sharing, external parties cannot assess any of the other four principles and the study is, in effect, a black box. These interactions do not form a fixed hierarchy. A study can be strong on one principle and weak on another simultaneously. The overall strength of a study's findings therefore depends on its full validity profile. Appendix~B operationalizes this by specifying for each principle what it does and does not license a researcher to conclude.

\subsection{From Principles to Guidelines}

Each principle is operationalized into a set of concrete, actionable guidelines documented in \hyperref[tab:guidelines]{Appendix~A}. Table~\ref{tab:schema} illustrates the structure of those guidelines: each entry specifies a short title, key implementation steps, the reasoning and evidence base behind the requirement, the research phase(s) at which it applies, and the validity principle(s) it addresses. The full set of 33 guidelines is presented in \hyperref[tab:guidelines]{Appendix~A}; readers designing a new study or reviewing an existing one can use that table as a checklist organized by each of these fields.

\begin{table*}[t]
\caption{Structure of the guidelines documented in \hyperref[tab:guidelines]{Appendix~A}. Each of the 33 guidelines is characterized across six fields. The illustrative example is drawn from Guideline~1.}
\label{tab:schema}
\vskip 0.1in
\begin{center}
\begin{small}
\setlength{\tabcolsep}{3pt}
\begin{tabular}{>{\RaggedRight}p{0.3cm}
                >{\RaggedRight}p{2.5cm}
                >{\RaggedRight}p{5.0cm}
                >{\RaggedRight}p{3.0cm}
                >{\RaggedRight}p{2.5cm}
                >{\RaggedRight}p{2.5cm}}
\toprule
\textbf{\#} &
\textbf{Guideline} &
\textbf{Key Implementation Points} &
\textbf{Rationale} &
\textbf{Research Phase} &
\textbf{Validity Principles} \\
\midrule
\multicolumn{6}{l}{\textit{Each field contains:}} \\[2pt]
1  &
Short descriptive title &
Operationalized steps, concrete checklists, and specific design and reporting choices &
Evidence base, reasoning, and consequences of non-adherence &
Design / Analysis / Reporting / Pre-Registration / Discussion / Post-publication &
One or more of: IV, EV, CV, SCV, TRV \\
\midrule
\multicolumn{6}{l}{\textit{Illustrative example (Guideline~1):}} \\[2pt]
1 &
Use evidentiary continuum for $p$-values &
Use $p{<}0.005$ for novel AI impact claims; report exact $p$-values with effect sizes and CIs; label strength as ``suggestive/strong/very strong'' rather than ``significant/non-significant'' &
Nascent field; false positives set bad precedents \citep{benjamin2018}
Analysis; Reporting &
SCV; TRV \\
\bottomrule
\end{tabular}
\end{small}
\end{center}
\vskip -0.1in
\end{table*}

\subsection{Extensions Beyond Prior Work}

Our framework extends prior work in several ways. On theory of change and failure modes, where RAND briefly mentions mechanisms and other frameworks focus on reporting or evaluation rigor, we explicitly articulate collecting data on mechanisms, cause and effect, linking AI outputs to human behavior, and listing plausible failure modes to guide design and interpretation. On heterogeneity, where prior work mentions stratification but lacks rigorous pre-specification and power planning, we pre-specify subgroups including baseline skill, AI literacy, demographics, and task characteristics, and we calculate power for interactions while distinguishing exploratory from confirmatory analyses. On AI-specific contamination and spillover, where prior work offers limited guidance on human-AI interaction contamination, we introduce spillover-aware designs, monitoring of AI usage, contamination prevention, and cluster randomization to handle real-world AI diffusion. On practical significance, where prior work emphasizes statistical significance with limited guidance on interpretation, we integrate practical significance, cost-benefit framing, equity and heterogeneity impact, ceiling and floor effects, sustainability, and domain-contextual interpretation. On evidentiary thresholds, where standard $\alpha = 0.05$ thresholds are common, we set $\alpha = 0.005$ for novel causal claims and emphasize effect size estimation, confidence intervals, and sensitivity to small-sample biases. On AI literacy stratification, where prior work defines population by domain or general demographics, we incorporate AI literacy stratification, explicit screening for AI experience, and non-WEIRD population sampling to improve external validity and heterogeneity analysis.

\section{Applications}

We intend for these principles and guidelines to function as shared infrastructure for the emerging field of AI evaluation RCTs. We see three main use cases.

\subsection{Design Aid for Researchers}

Researchers planning new AI evaluation RCTs can use the guidelines as a pre-study checklist, prompting explicit decisions about the unit of randomization, contamination controls, heterogeneity analysis, evidentiary thresholds, transparency commitments, and other methodological choices. The guidelines operationalize abstract principles into concrete implementation steps accessible to engineers and applied scientists without extensive RCT training.

\subsection{Evaluation Rubric for Reviewers and Institutions}

Funders, journals, policymakers, and internal governance bodies can use the framework as a review rubric. There is a broad need for this kind of structured review. For example, a funder assessing the state of evidence on a policy question, a journal reviewer evaluating a submitted manuscript, a policymaker trying to determine whether a body of research is sufficient to act on, or an AI governance body trying to assess whether frontier labs are meeting something like the requirements of the EU AI Act all face a shared problem. The problem is that existing AI evaluation research is often reported in ways that make it difficult to determine what was actually done, what the findings warrant, and how findings compare across studies. Standardized review criteria address this directly.

In practice, applying the framework as a rubric involves working through its components in sequence. A reviewer can begin with the five validity principles as an interpretive lens. For each principle, what does the paper claim, and does the study design support that claim? Appendix B's inferential scope table operationalizes this check by specifying, for each principle, what a study can and cannot conclude. Appendix B functions as a structured prompt for identifying gaps between a paper's stated claims and what its design actually enables. The reporting template provides a parallel check at the level of the findings themselves. A well-scoped claim should be expressible in the form "In [population/task], assignment to [AI access condition] caused an estimated change of [effect size and confidence interval] in [human outcome], compared with [counterfactual], over [study period]; generalization beyond [scope] requires additional evidence." Claims that cannot be translated into this form typically signal an inferential overreach.

The 33 guidelines in Appendix A support a more systematic review. Each guideline specifies the research phase it applies to and the validity principle it addresses, so a reviewer can identify which phases are underreported and which validity dimensions receive no attention. A paper that documents design and analysis choices but omits pre-registration and disclosure practices, for example, fails on TRV even if its causal identification is otherwise sound.

The glossary serves a diagnostic function as well. Terminological imprecision (e.g. papers that use "uplift" to mean treatment effect, effect size, or capability gain interchangeably, or that conflate reproducibility with replicability) is often a signal of underlying conceptual confusion about what the study is actually estimating. Reviewers can use the glossary both to interpret papers written under different terminological conventions and to flag cases where imprecision in language reflects imprecision in design or inference.

Taken together, these materials give reviewers a reusable template: work through the validity principles, check inferential scope against Appendix B, evaluate claim language against the reporting template, assess phase and principle coverage using Appendix A, and use the glossary to interpret or flag terminological choices. The more consistently reviewers apply the same criteria across papers, the more useful individual reviews become as inputs to the broader project of understanding what the field actually knows.

\subsection{Field-Level Standard Setting}

The principles and guidelines can act as a roadmap for field-level standard setting. This paper's aim is to provide conceptual and operational foundations for that process rather than to complete it. The detailed guideline list can be selectively elevated into formal requirements (e.g. mandatory pre-registration or minimum reporting items) as the field matures and institutions converge on norms.

Governance and standard-setting uses of this framework should treat it as a structured starting point pending further development, not a finalized standard. This is consistent with how analogous frameworks positioned themselves early on. The TOP Guidelines offer one example of this trajectory. They were originally proposed by a cross-disciplinary committee in 2015, TOP was refined over a decade of adoption across thousands of journals and organizations, with a formal advisory board and public comment process producing a substantially updated version in 2025. This paper is intended as a contribution to an analogous process for AI evaluation RCTs, not its conclusion.

Together, these three use cases point toward the kind of credible, cumulative knowledge production that evidence-based AI policy and safety frameworks require.

\section{Limitations and Future Work}

Our framework focuses exclusively on experimental designs for assessing causal effects of AI systems on human outcomes. This deliberate scope limitation means we do not address observational studies, purely benchmark-based evaluations, or quasi-experimental designs that may be more feasible in certain organizational contexts. The comprehensive nature of our guidelines, particularly requirements for pre-registration, multi-modal data collection, heterogeneity analysis with adequate statistical power, and tiered transparency frameworks, imposes demands some research teams may find challenging. A tiered or partial-adoption version of the framework would appear to address this, but the guidelines are highly interdependent. The guidelines address distinct validity dimensions, each of which must hold for a study's inferential claims to be credible. Gaps in one do not average out against strengths in another. Heterogeneity analysis, for instance, is underpowered without adequate sample size. Pre-registration without outcome disclosure provides weak protection against selective reporting. A path taken by analogous frameworks in other fields is to invest in implementation infrastructure that lowers the cost of meeting the standard. This could include pre-registration templates, statistical power calculators tailored to AI studies, and training resources for teams without deep experimental methods expertise. Adoption is also a process of culture change that runs from practices being possible, to easy, to expected, to rewarding, to required, as the field builds capacity and institutions align around shared norms. The TOP Guidelines followed something close to this arc over a decade, and a realistic expectation for this framework is similar.

AI systems evolve at a rate that outpaces traditional experimental timeframes. Our guidelines address model versioning and drift monitoring, but the fundamental challenge of evaluating rapidly changing systems persists. As AI capabilities advance toward more agentic and autonomous systems, human-AI interaction dynamics may shift in ways our current frameworks cannot fully anticipate.

This framework has not yet been validated through systematic community review, pilot implementation, or empirical assessment of whether adherence actually improves research quality. The framework draws on methodological traditions, particularly the Shadish-Cook-Campbell validity structure and open-science practices developed across medicine, psychology, economics, and related fields, that have accumulated substantial scrutiny and refinement over decades. What lacks this prior scrutiny is how these components work in combination, and how well the AI-specific adaptations hold up in practice. These are the aspects most in need of expert review and pilot testing before the framework is used to design studies, evaluate existing research, or inform governance, funding, and policy decisions. Future work should pursue this through structured engagement with researchers from fields with deep experience in experimental methods, following processes analogous to those that produced other frameworks and guidelines for research practices. It should also consider developing implementation tools such as pre-registration templates and statistical power calculators tailored to AI studies, and create domain-specific adaptations of the core framework.

\section{Conclusion}

The advancement of AI capabilities demands evaluation practices that are both rigorous and responsive to the unique challenges these systems present. This work establishes a cross-disciplinary framework for standardizing AI evaluation research, synthesizing established experimental practices into actionable guidance structured around five core principles: construct validity, internal validity, external validity, statistical conclusion validity, and transparency, repeatability, and verification.

We extend prior reporting standards by formalizing causal reasoning through RCT methodology adapted for AI's dynamic nature, integrating heterogeneity analysis with adequate statistical power, implementing stringent evidentiary thresholds, establishing a graded transparency framework, and addressing AI-specific challenges including model versioning, contamination effects, and equitable impact assessment. Rigorous, transparent, and repeatable evaluation practices are essential safeguards ensuring that AI development proceeds on credible evidence, that benefits are distributed equitably, that risks are identified proactively, and that claims about AI capabilities align with empirical reality.

\section*{Impact Statement}

This paper presents work whose goal is to advance the field of AI Safety. By establishing rigorous standards for evaluating how AI systems affect human performance, we aim to improve the credibility of evidence informing deployment decisions, safety assessments, and governance policies. There are many potential societal consequences of our work, none which we feel must be specifically highlighted here.

\section*{Use of Large Language Models}
Large language models (Claude, Anthropic) were used during the development of this manuscript and associated materials. Based on resources, drafts, and ideas presented by the authors, LLMs were used for brainstorming and suggesting candidate principles and guidelines that were then curated and revised by authors. Additional uses include LaTeX formatting and conference poster preparation. All substantive intellectual judgments (e.g. including the selection, framing, and argumentation for the final set of principles, the theoretical framework, and normative positions) reflect the authors' own analysis, with additional input from external reviewers and collaborators.

\section*{Acknowledgements}

This work was carried out through the Supervised Program for Alignment Research (SPAR), whose program structure and infrastructure support made the research possible. We thank Ayrton San Joaquin for helpful comments on an earlier draft. All errors remain our own.

%\nocite{langley00}
\bibliography{references}
\bibliographystyle{icml2026}

% =====================================================================
% GUIDELINES TABLE
% =====================================================================
\newpage
\onecolumn

\begin{landscape}

\small
\setlength{\LTcapwidth}{\linewidth}
\setlength{\tabcolsep}{4pt}

\begin{longtable}{%
  >{\RaggedRight}p{0.4cm}%   #
  >{\RaggedRight}p{3.0cm}%   Guideline
  >{\RaggedRight}p{7.5cm}%   Key Implementation Points
  >{\RaggedRight}p{3.5cm}%   Rationale (brief)
  >{\RaggedRight}p{3.0cm}%   Research Phase
  >{\RaggedRight}p{3.5cm}%   Validity Principles
}

\caption*{\textbf{Appendix A.} Methodological Guidelines for AI evaluation RCTs. Abbreviations: SCV = Statistical Conclusion Validity; IV = Internal Validity; CV = Construct Validity; EV = External Validity; TRV = Transparency, Repeatability \& Verification.}
\label{tab:guidelines}\\

\toprule
\textbf{\#} & \textbf{Guideline} & \textbf{Key Implementation Points} & \textbf{Rationale} & \textbf{Research Phase} & \textbf{Validity Principles} \\
\midrule
\endfirsthead

\multicolumn{6}{l}{\small\itshape Appendix A continued from previous page.}\\
\toprule
\textbf{\#} & \textbf{Guideline} & \textbf{Key Implementation Points} & \textbf{Rationale} & \textbf{Research Phase} & \textbf{Validity Principles} \\
\midrule
\endhead

\midrule
\multicolumn{6}{r}{\small\itshape Continued on next page.}\\
\endfoot

\bottomrule
\endlastfoot

%--- Guideline 1 ---
1 & Use an evidentiary continuum to interpret statistical significance &
Interpret $p$-values as continuous evidence, not binary pass/fail. For novel AI causal claims, use $p < 0.005$ as threshold \citep{benjamin2018}. Report exact $p$-values with effect sizes and CIs. Label strength as ``suggestive/strong/very strong'' rather than ``significant/non-significant.'' Less stringent thresholds may apply for direct replications or very large pre-registered samples ($n > 1000$/group). &
Early AI evaluations studies can set field precedents. Making these rigorous is critical in a nascent field with many researcher degrees of freedom and high-stakes deployment implications. &
Analysis; Reporting &
SCV; TRV \\

\midrule

%--- Guideline 2 ---
2 & Report effect sizes with confidence intervals, not just $p$-values &
Always report: (1) effect size in original units; (2) standardized effect size (Cohen's $d$, Hedges' $g$, odds ratios); (3) 95\% CIs; (4) $p$-values. Additionally report practical significance thresholds (pre-specified), cost-benefit framing, and heterogeneity metrics (range of individual effects, $\tau^2$). &
$p$-values alone do not convey magnitude or practical importance. Effect sizes are required for cost-benefit analysis, meta-analysis, and deployment decisions. &
Analysis; Reporting &
TRV \\

\midrule

%--- Guideline 3 ---
3 & Distinguish statistical significance, practical significance, and scientific importance &
Explicitly discuss all three types of significance. Contextualize effect sizes relative to: domain baseline, implementation costs, scale of application, task criticality, and time horizon. Pre-specify practical significance thresholds before analysis. Avoid applying Cohen's conventions without domain justification. &
The same numerical effect has vastly different implications across domains. Decision-makers need magnitude and cost-effectiveness information, not just detectability. &
Design; Analysis; Reporting &
SCV; TRV \\

\midrule

%--- Guideline 4 ---
4 & Pre-register the study protocol and transparently report adherence &
Register prospectively (OSF, AsPredicted, ClinicalTrials.gov, AEA RCT Registry). Log registry URL and date. Make full protocol and statistical analysis plan publicly accessible with version history. Report adherence explicitly; distinguish confirmatory from exploratory analyses; document and justify any deviations in real-time. Share completed CONSORT-AI or SPIRIT-AI checklist as supplementary material. &
Pre-registration only improves credibility if researchers adhere to the plan. Transparent deviation reporting maintains scientific integrity while acknowledging unforeseen issues. &
Design; Analysis; Reporting; Pre-Registration &
IV; SCV; TRV \\

\midrule

%--- Guideline 5 ---
5 & Design and report studies to enable systematic aggregation &
Use consistent effect size metrics and sufficient statistics for meta-analysis (means, SDs, $n$ per group, exact $p$-values). Follow CONSORT reporting standards. Pre-specify and report key moderators. Report variance of individual effects ($\tau^2$), intraclass correlations for clustered designs, and subgroup analyses with interaction tests. Plan and execute participant-level data sharing (OSF, Dataverse). Present CONSORT-style participant flow diagram. &
Individual AI evaluation studies provide limited evidence. Standardized reporting and data sharing enable meta-analysis, bias correction, and evidence synthesis essential for deployment decisions. &
Design; Analysis; Reporting; Post-publication &
EV; SCV; TRV \\

\midrule

%--- Guideline 6 ---
6 & Assess evidence sufficiency, provide contextualized interpretation, and report generalizability &
Define target population and distinguish it from the statistical sample. Systematically evaluate external validity threats (selection, setting, temporal, AI system specificity). Document boundary conditions; state when findings should NOT be extrapolated. Flag evidence gaps and required follow-up studies. Contextualize findings relative to prior evidence; avoid overstatement. &
AI evaluation RCT findings guide high-stakes deployment decisions. Evidence-to-policy gaps are large when convenience samples and controlled settings are extrapolated to production environments. &
Design; Reporting; Discussion &
EV; CV; TRV \\

\midrule

%--- Guideline 7 ---
7 & Communicate and disseminate evidence in accessible, policy-relevant formats &
Structure findings to answer: effectiveness questions (for whom, under what conditions), heterogeneity questions, implementation requirements, boundary conditions, and evidence gaps. Formats: structured abstracts, GRADE-style evidence profiles, plain language summaries, evidence briefs. Disseminate via preprint, open-access publication, trial registry update, conference presentations spanning AI, domain, and policy venues. &
Decision-makers without methodological training need structured guidance on whether AI evidence applies to their context. Multi-channel dissemination maximizes real-world impact. &
Reporting; Dissemination; Discussion &
EV; CV; TRV \\

\midrule

%--- Guideline 8 ---
8 & Plan for heterogeneity analysis across meaningful subgroups &
Pre-specify subgroups and outcomes. Focus on theoretically meaningful groups: baseline skill/expertise, demographics, task characteristics, AI usage patterns, and temporal learning effects. Use interaction terms (not subgroup-specific tests). Ensure adequate power for planned subgroup analyses. Clearly distinguish confirmatory from exploratory subgroup analyses. &
AI likely affects different people differently. Mean effects can mask harm to subgroups. Pre-specification prevents inflated false positive rates from exploratory subgroup mining. &
Design; Analysis &
EV \\

\midrule

%--- Guideline 9 ---
9 & Determine appropriate sample size through power analysis ($\alpha \leq 0.005$) &
Identify primary research question and inferential goals before calculating power. Plan conservatively: use $d = 0.3$--$0.4$ as target effect size in absence of field benchmarks. Note that testing interactions requires $\approx$4$\times$ the sample size of main effects. If underpowered for subgroup analysis, state this explicitly. Consider equivalence testing for null/small effect hypotheses. &
AI evaluation is a nascent field with limited prior effect size evidence and risk of winner's curse inflation. Stringent $\alpha$ thresholds increase required sample sizes; account for this in design. &
Design; Analysis &
IV; EV; SCV \\

\midrule

%--- Guideline 10 ---
10 & Specify population clearly &
Define the subset of humans the study represents using relevant dimensions: domain expertise and AI literacy. Justify why this population was chosen for the research question. Report how well the achieved sample compares to the target population, both overall and by treatment arm. &
Vague population definitions (e.g., ``users,'' ``workers'') prevent generalizability assessment. AI effects vary substantially by expertise and prior AI exposure; sample composition is critical for interpretation. &
Design; Reporting; Discussion &
EV; CV \\

\midrule

%--- Guideline 11 ---
11 & Implement participant quality control &
Apply pre-specified inclusion/exclusion criteria: domain knowledge tests, attention checks, language proficiency verification, technical setup verification, performance-based thresholds. Post-hoc exclusions must be pre-registered. Report all exclusions with counts and reasons. Conduct sensitivity analyses with and without exclusions. &
Inattentive or unqualified participants introduce noise, reduce power, and can bias effect estimates. AI tools may mask participant quality, making QC especially critical in RCT studies. Eligibility criteria are pre-specified in the participant eligibility guideline; this guideline focuses on implementation-stage 
  quality control.                &
Design; Analysis; Reporting &
IV; SCV; EV \\

\midrule

%--- Guideline 12 ---
12 & Stratify randomization on relevant variables and fully document allocation procedures &
Measure AI literacy and domain expertise before randomization. For $n < 500$, use blocked randomization within strata. Document: randomization software, random seed, timing relative to enrollment, allocation ratio, who performed randomization, and allocation concealment mechanisms. Verify and report balance across groups post-randomization. &
Stratification ensures comparable expertise distributions across arms and enables adequately powered subgroup analyses. Inadequate allocation documentation is a leading source of bias and a common CONSORT reporting deficiency. &
Design; Analysis; Reporting &
IV; EV; SCV; TRV \\

\midrule

%--- Guideline 13 ---
13 & Monitor compliance and track actual treatment exposure &
Implement objective tracking: API restrictions, browser extensions, activity logs, screen recordings. Define compliance operationally (e.g., assigned AI tool used for $>$80\% of tasks). Track partial compliance continuously. Conduct intention-to-treat (ITT) analysis as primary; complier-average causal effect (CACE) as sensitivity check. Assess differential non-compliance across conditions. &
Non-compliance threatens internal validity via discrepancies between assigned and received treatment. AI tools are freely available, making control contamination likely in online studies. &
Design; Analysis; Reporting &
IV; SCV; EV \\

\midrule

%--- Guideline 14 ---
14 & Ensure UI equivalence across conditions &
Design interfaces that are identical across treatment and control groups except for AI assistance. Provide a ``placeholder'' interface for control (same layout, AI features disabled). Match cognitive load, information access, and interaction complexity. Avoid bundling AI with superior interface features. Pilot-test both interfaces for equivalence. &
UI differences confound treatment effects with interface effects, preventing attribution of observed differences to AI capability alone. &
Design; Analysis &
CV; IV \\

\midrule

%--- Guideline 15 ---
15 & Control for and measure participant effort &
Implement interaction logging: timestamps for task start/completion, active vs.\ idle time, AI query counts, revision iterations, mouse/keyboard activity. Conduct embedded attention checks. Report outcomes with and without time controls. Examine effort-outcome tradeoffs (speed vs.\ quality). Verify manipulation success (treatment used AI; control did not). &
Observed performance differences could reflect differential effort rather than capability differences. AI access may increase motivation and engagement, confounding productivity estimates. &
Design; Analysis; Reporting &
IV; CV \\

\midrule

%--- Guideline 16 ---
16 & Implement blinding and ensure consistent, reliable outcome assessment &
Report blinding status for all roles: participants, researchers/administrators, outcome assessors, and analysts. When blinding is impossible (AI presence is obvious), state this explicitly. For human-graded outputs: remove metadata, randomize output order, use blinded graders not involved in other study aspects. Calculate inter-rater reliability (ICC, Krippendorff's $\alpha$, Cohen's $\kappa$); set minimum thresholds (e.g., ICC $> 0.70$). Conduct calibration sessions and monitor rater drift. &
Knowledge of treatment assignment biases behavior, assessment, and analysis at every study stage. Inconsistent grading reduces power and can introduce condition-correlated bias. &
Design; Analysis; Reporting &
IV; CV; SCV \\

\midrule

%--- Guideline 17 ---
17 & Collect qualitative data to understand mechanisms &
Complement quantitative outcomes with: think-aloud protocols, screen recordings, chat logs, semi-structured post-task interviews, open-ended surveys, and perception measures. Link qualitative themes to quantitative patterns. Use qualitative data to explain heterogeneity and identify unexpected mechanisms. Generate hypotheses for future quantitative testing. &
Quantitative outcomes answer ``what happened''; qualitative data answers ``how and why.'' Mechanistic understanding is essential for generalizing beyond specific tool-task-population combinations. &
Design; Analysis; Reporting &
CV; IV; EV \\

\midrule

%--- Guideline 18 ---
18 & Ensure metric consistency across conditions &
Apply identical quality rubrics, performance benchmarks, and scoring criteria to human-only and human+AI outputs. Pre-register all evaluation criteria before data collection. Pilot-test metrics on sample outputs. Prohibit mid-study metric changes; if necessary, report both original and revised analyses. Avoid post-hoc metric exploration when initial metrics show null results. &
Inconsistent metrics confound treatment effects with measurement differences. AI may excel on some quality dimensions differently than humans; temptation to redefine ``quality'' post-hoc must be resisted. &
Design; Analysis; Reporting &
CV; IV \\

\midrule

%--- Guideline 19 ---
19 & State study design, population, intervention, framework, and phase in title and abstract &
Title must specify: study design (e.g., parallel-group RCT), population (e.g., ``novice programmers''), AI tool, and comparison condition. Abstract must state conceptual framework (superiority, non-inferiority, equivalence, or exploratory) and trial phase (proof-of-concept, efficacy, effectiveness, or implementation). Avoid vague language (e.g., ``we examine AI's effects''). &
Descriptive titles and abstracts enable rapid relevance assessment, evidence synthesis, and detection of outcome switching or protocol deviations. Framework statements prevent misinterpretation of exploratory findings as confirmatory evidence. &
Design; Reporting; Pre-Registration &
SCV; CV; TRV \\

\midrule

%--- Guideline 20 ---
20 & Specify eligibility criteria, key inclusion/exclusion criteria, and study setting &
State population definition with relevant dimensions (expertise, AI literacy). List inclusion criteria (qualification thresholds, technical requirements, availability) and exclusion criteria (prior exposure, conflicts of interest, technical limitations). Document setting: physical/virtual location, supervision level, environmental conditions, technical infrastructure, timing. Define ``screened,'' ``enrolled,'' ``completed,'' and ``analyzed.'' Report AI literacy stratification approach. &
Vague eligibility descriptions make it impossible to assess sample representativeness, compare across studies, or determine boundary conditions. Setting characteristics critically affect external validity. Implementation-stage quality control procedures for enrolled participants are addressed in the participant quality control guideline. &
Design; Reporting &
EV; TRV; IV \\

\midrule

%--- Guideline 21 ---
21 & Report important harms, unintended effects, and adverse events &
Pre-specify harm categories: skill atrophy, automation complacency, over-reliance, de-skilling, workflow disruption, equity effects, psychological impacts. Collect harm data for both conditions using identical instruments. Report all harms by group (frequency, severity, duration), not only statistically significant ones. Discuss risk-benefit balance in main results (not supplementary). Monitor for differential dropout as harm signal. &
Publication bias already favors positive findings; selective non-reporting of harms compounds this. AI harms may be cognitive/psychological and emerge only after sustained use or tool removal. &
Design; Analysis; Reporting &
IV; EV; CV; TRV \\

\midrule

%--- Guideline 22 ---
22 & Share data, materials, and code with full version control and documentation &
Deposit de-identified participant-level data in public repositories (OSF, Dataverse, Zenodo) with DOIs. Share complete analysis pipeline (raw data $\to$ outputs) with comprehensive codebook. Share all study materials: instructions, rubrics, AI prompts. Use Git for version control with descriptive commits; tag releases; include commit hashes in published results. Provide a clear README and license (e.g., CC-BY for data, MIT for code). &
Shared data and code enable independent verification, re-analysis, and individual participant data meta-analysis. Version control prevents the common problem of unreproducible analyses months after completion. &
Design; Reporting; Post-publication &
TRV; EV; SCV \\

\midrule

%--- Guideline 23 ---
23 & Document roles, responsibilities, funding sources, and conflicts of interest &
Identify all contributors using CRediT taxonomy. Document who held decision-making authority over design, analysis, and publication. Report all monetary and in-kind support (including AI company API access). Disclose financial and non-financial COI for all investigators and committee members. Explicitly flag any support from AI companies whose products are under evaluation. Use ICMJE COI disclosure forms. &
Studies supported by AI companies with commercial stakes in capability claims face inherent conflicts. Standardized disclosure enables readers to assess potential for subtle influence on task selection, outcome measures, and reporting. &
Reporting; Pre-Registration &
TRV; IV \\

\midrule

%--- Guideline 24 ---
24 & Articulate theory of change, hypotheses, assumptions, and failure modes before data collection &
State the specific causal mechanism (e.g., AI reduces cognitive load, compensates for expertise gaps, accelerates feedback). Map causal pathway: AI feature $\to$ user interaction $\to$ intermediate process $\to$ outcome. List all assumptions (population, task, AI system, measurement). State all hypotheses with expected effect direction and size. Pre-specify plausible failure modes (non-adoption, error propagation, effort reduction, task mismatch) with detection criteria. &
Without a theory of change, positive results are unexplained, null results are uninterpretable, and replication is directionless. Multiple competing mechanisms may explain any given result. &
Design; Pre-Registration &
CV; IV; EV \\

\midrule

%--- Guideline 25 ---
25 & Establish causality through RCT design with carefully controlled conditions &
Use RCTs as gold standard. Ensure conditions are identical except for AI access. Specify and justify control condition (no tool, placebo interface, alternative tool, or standard practice). Identify and address AI-specific confounds: Hawthorne effects, novelty effects, differential motivation. Avoid bundling AI with other changes (better UI, additional training). When RCTs are infeasible, use strongest available quasi-experimental design (RD, DiD, IV) and explicitly discuss assumptions required. &
Voluntary AI adoption creates severe selection bias in observational studies. Random assignment is the only credible method for estimating the causal effect of making AI available. &
Design &
IV; CV; SCV \\

\midrule

%--- Guideline 26 ---
26 & Define the Overall Evaluation Criterion (OEC) before the experiment begins &
Pre-specify a single primary outcome (or pre-registered composite with justified weighting) representing what ``success'' means. Distinguish primary outcomes (confirmatory, adequately powered) from secondary outcomes (potentially exploratory). Align OEC with deployment decision type. Document minimum detectable effect and practical significance threshold tied to OEC. Do not change primary outcome after seeing data; if necessary, report both original and modified analyses. &
Without a pre-specified OEC, researchers face large temptation to explore multiple outcomes and report selectively. Pre-committed composites force explicit tradeoff specification before results are known. &
Design; Pre-Registration &
SCV; CV; IV \\

\midrule

%--- Guideline 27 ---
27 & Refine measurement instruments through iterative validation, pre-testing, and piloting &
Conduct cognitive interviews/think-alouds to verify participant comprehension of instructions and survey items. Pilot full protocol with a non-main-study sample. Assess inter-rater reliability, test-retest reliability, and internal consistency. Validate rubrics against expert judgment. Check for ceiling/floor effects. Test full technical infrastructure (randomization system, AI integration, timing). Report all pilot findings and resulting protocol modifications. &
Standardized instruments for measuring AI's impact on human performance largely do not exist; bespoke measures require validation. AI tool integration introduces technical failure modes detectable only through piloting. &
Design; Pre-Registration &
CV; SCV; IV \\

\midrule

%--- Guideline 28 ---
28 & Satisfy ethical requirements including informed consent appropriate for digital studies &
Obtain and report IRB/REB compliance. Adhere to core principles: Respect for Persons, Beneficence, Justice. Inform participants about AI involvement, data processing, and interaction logging. Use electronic consent with affirmative action (no pre-checked boxes). Allow granular consent for different data uses. Include comprehension checks. Address ethical reciprocity (offer AI access to control group post-study). If deception is used, obtain ethics board approval and conduct debriefing. &
Online and remote studies face unique consent challenges; AI-specific elements require explicit disclosure. Ethics review and informed consent are foundational to RCT validity and participant protection. &
Design; Reporting; Pre-Registration &
TRV; IV; EV \\

\midrule

%--- Guideline 29 ---
29 & Conduct and report robustness checks and sensitivity analyses &
Check robustness to: alternative model specifications, different covariate adjustments, alternative variable measures, different subsamples (outlier exclusion), missing data handling, and functional form. Examine sensitivity to assumption violations, influential observations, timing of effects, AI usage heterogeneity, compliance variation, learning effects, task order effects, and AI system changes during study. Report all sensitivity analyses; distinguish pre-specified from post-hoc. &
Every analysis involves judgment calls. Findings robust across specifications inspire greater confidence. Transparency about sensitivity analyses helps identify boundary conditions and prevents motivated reporting. &
Design; Reporting &
IV; SCV; TRV \\

\midrule

%--- Guideline 30 ---
30 & Stipulate procedures for linking baseline, treatment, and outcome data &
Assign unique participant IDs at enrollment. Use consistent ID format across all data sources and timepoints. Create data dictionary mapping IDs across files. Pre-specify and document linkage procedures; test with pilot data. Implement validation checks for linkage accuracy. Document de-identification procedures. Store linking documentation separately from analysis code. Pre-register linking methodology. &
Data linkage failures result in loss of data, incorrect pairing of covariates with outcomes, and attribution errors that invalidate analyses. Linkage documentation is essential for independent verification, especially in multi-site or multi-wave studies. &
Design; Analysis &
IV; TRV \\

\midrule

%--- Guideline 31 ---
31 & Develop comprehensive participant management, communication, and data handling protocols &
Pre-study: define protocol violations with examples; pre-register exclusion criteria; develop SOPs and decision flowcharts. Screening: use structured interviews with yes/no decision points; track screening-to-enrollment ratios. Communication: designate contact person, establish response timeframes (e.g., 24-48 hours), create FAQ and response templates, document all communications. Withdrawal: create clear withdrawal process; conduct exit interviews; analyze patterns by condition. Violations: pre-specify when violations trigger exclusion; use blinded review. Prepare CONSORT flow diagram. &
Flexible exclusion rules and undocumented decisions are primary drivers of false positives and replication failures. Standardized protocols prevent p-hacking through pre-specification and create audit trails for verification. &
Design; Analysis; Reporting &
IV; SCV; TRV; EV \\

\midrule

%--- Guideline 32 ---
32 & Monitor data quality and critical study metrics in real-time with pre-specified stopping rules &
Build a dashboard tracking: enrollment rate, completion rate, balance across conditions, attrition by condition, data quality flags (fast completions, identical responses, straight-lining, duplicate IPs), and treatment exposure rates. Define pre-specified quantitative thresholds triggering review, enrollment pause, or data collection halt. Document all data quality decisions in a timestamped log. Report quality metrics and any actions in the final manuscript. &
Real-time monitoring enables early detection and correction of data quality problems. AI tool access may change participant behavior in unexpected ways detectable only through active monitoring. Pre-specified stopping rules prevent both premature termination and continued collection through known problems. &
Design; Analysis; Reporting &
IV; SCV; TRV \\

\midrule

%--- Guideline 33 ---
33 & Determine what baseline data to collect and collect it before randomization &
Collect before randomization or treatment assignment disclosure: demographics (age, education, experience), AI-specific measures (prior tool usage, AI literacy, attitudes), domain competence (self-reported and objective pre-task performance), psychological measures (self-efficacy, motivation), and technical setup. Use baseline task performance as covariate and potential stratification variable. Time-stamp all baseline data collection. &
Baseline data collected after randomization introduces bias. AI literacy and domain expertise are the strongest expected moderators of AI effects; failing to measure them means missing the most important heterogeneity in the data. &
Design; Pre-Registration &
IV; EV; SCV \\

\end{longtable}

\end{landscape}

% =====================================================================
% APPENDIX TABLE B: INFERENTIAL SCOPE BY VALIDITY PRINCIPLE
% =====================================================================

\clearpage
\onecolumn
\begin{landscape}

\footnotesize
\setlength{\LTcapwidth}{15.6cm}
\setlength{\tabcolsep}{3pt}

\definecolor{principlebg}{gray}{0.80}

\begin{longtable}{%
  >{\RaggedRight}p{2.8cm}%   Validity Principle
  >{\RaggedRight}p{4.4cm}%   Enables you to conclude
  >{\RaggedRight}p{4.4cm}%   Does NOT enable you to conclude
  >{\RaggedRight}p{4.0cm}%   Appropriate language
}

\caption*{\textbf{Appendix B.} Inferential Scope by Validity Principle. This table specifies what each validity principle does and does not license researchers to conclude, along with example language. Validity principles are drawn from the Campbell framework \citep{shadish2002}, extended with a fifth principle for transparency, repeatability, and verification (TRV). \textit{Repeatability} is used as an umbrella term for reproducibility, robustness, and replicability following \citet{nosek2026glossary}; see the TRV principle definition for full terminology notes. The five principles are interdependent: the validity of the overall evidence package depends on all five dimensions considered together.}
\label{tab:inferential}\\

\toprule
\textbf{Validity Principle} & \textbf{Enables you to conclude} & \textbf{Does NOT enable you to conclude} & \textbf{Appropriate language} \\
\midrule
\endfirsthead

\multicolumn{4}{l}{\small\itshape Appendix B continued from previous page.}\\
\toprule
\textbf{Validity Principle} & \textbf{Enables you to conclude} & \textbf{Does NOT enable you to conclude} & \textbf{Appropriate language} \\
\midrule
\endhead

\midrule
\multicolumn{4}{r}{\small\itshape Continued on next page.}\\
\endfoot

\bottomrule
\endlastfoot

% ------------------------------------------------------------------
% CV ROW
% ------------------------------------------------------------------
\textbf{Construct Validity (CV)}

\textit{Are we measuring what we claim to measure?} &

The measures used validly represent the theoretical constructs of interest (e.g., decision quality, coding correctness, clinical reasoning), so estimated effects pertain to those constructs rather than to artifacts, proxies, or construct-irrelevant variance. Observed changes reflect genuine differences in the intended capability or performance construct, not merely in surface metrics such as speed, token count, or verbosity. &

Causation (that is the role of internal validity) or generalization to other contexts (external validity). That other operationalizations of the same construct would behave identically without additional validation. That measured gains reflect specific mechanisms (e.g., genuine reasoning improvement vs.\ template reuse) absent targeted mediator tests. That construct validity in one condition guarantees measurement equivalence across conditions --- invariance testing is required. &

``Our outcome operationalization is a valid measure of X (evidence: content, criterion, and/or convergent-discriminant validation).'' \newline
``We interpret effects as changes in X, not merely in proxy Y (e.g., tokens, lines of code, time-on-task).'' \newline
``Results hinge on this operationalization; alternative measures of X may yield different magnitudes or directions.'' \\

\midrule

% ------------------------------------------------------------------
% IV ROW
% ------------------------------------------------------------------
\textbf{Internal Validity (IV)}

\textit{Did AI access cause the difference?} &

The observed difference in outcomes was caused by the AI intervention as defined (e.g., assignment to access, or access to a pinned model version) for the study units over the study period. Alternative explanations --- including selection, history, maturation, instrumentation, contamination, and interference (if addressed by design) --- are unlikely to account for the observed effect. ITT effects reflect the impact of being assigned AI access; TOT effects reflect the impact among those who actually used it; these estimands are distinct and should be reported separately. &

Generalization to other populations, tasks, organizations, or AI system versions (that is the role of external validity). The mechanisms through which AI produced the effect unless directly tested with mediator designs or process measures. Effects for participants who never complied with their assigned condition, or under different treatment definitions (e.g., different model versions, temperatures) than those studied. That the effect will persist beyond the study period. &

``Assignment to AI assistance caused an average change of $\Delta$ (ITT estimate, 95\% CI [lower, upper]).'' \newline
``For participants who actually received and used AI, the estimated causal effect is [TOT estimate].'' \newline
Avoid: ``AI causes improvement in X'' without specifying the treatment definition, estimand, and study period. \\

\midrule

% ------------------------------------------------------------------
% EV ROW
% ------------------------------------------------------------------
\textbf{External Validity (EV)}

\textit{To whom and where do findings generalize?} &

The estimated effects generalize to explicitly specified target populations, tasks, time frames, settings, and AI implementations for which the study's sampling and transport assumptions plausibly hold. Findings are robust across the meaningful variations that were prospectively planned or replicated (e.g., across user expertise bands, across related model families, across study sites). The inferential scope is explicit: results apply to [defined target context] under [stated conditions]. &

Universal applicability across all users, tasks, organizations, or future model generations --- especially amid rapid model drift. Causation in the absence of internal validity. Transportability to contexts with substantially different constraints (e.g., different oversight intensity, incentive structures, risk levels) without additional evidence or explicit transport analysis. That findings from convenience or WEIRD samples apply to underrepresented populations without replication. &

``Findings are likely to generalize to [defined target context], given our sampling frame, replication across [sites/conditions], and model family checks.'' \newline
``Generalization to [new context, model, or version] is uncertain and requires replication or transport analyses.'' \newline
``Effects may differ for [subgroups or contexts not studied]; findings should not be interpreted outside their stated scope.'' \\

\midrule

% ------------------------------------------------------------------
% SCV ROW
% ------------------------------------------------------------------
\textbf{Statistical Conclusion Validity (SCV)}

\textit{Is there really an effect, how large is it, and how certain are we?} &

There is (or is not) an effect, with quantified uncertainty appropriate to the design --- accounting for clustering, repeated measures, and interference-aware estimands where applicable. The magnitude and direction of the effect are estimated with acceptable precision; error rates are controlled given multiplicity and researcher degrees of freedom. Statistical model assumptions are adequately met or robustly addressed. Effect sizes and confidence intervals are reported alongside $p$-values. &

What the effect means in theoretical terms (construct validity) or whether it generalizes (external validity). That statistically significant effects are practically important --- or that non-significant results imply no effect --- without power and precision context. Mechanisms, unless the analysis identifies them with appropriate design elements. That a small $p$-value indicates a large or practically meaningful effect, or that a large $p$-value indicates the absence of an effect. &

``Estimated average effect $= \Delta$, 95\% CI [lower, upper]; the analysis is powered for MDE $=$ [value] under [design assumptions].'' \newline
``Results are robust to alternative specifications, cluster-robust SEs, and pre-specified multiplicity adjustments.'' \newline
``We emphasize effect sizes and CIs over binary significance; practical importance is addressed separately.'' \\

\midrule

% ------------------------------------------------------------------
% TRV ROW
% ------------------------------------------------------------------
\textbf{Transparency, Repeatability \& Verification (TRV)}

\textit{Can others see what was done, verify it was done, and check whether findings hold?} &

The study's procedures, data transformations, code, and AI system details are sufficiently disclosed (Level~1) and accessible (Level~2) to permit critical appraisal and independent reuse. What was claimed to be shared has been independently verified as genuinely accessible and complete (Level~3). Reported results survive at least one form of independent repeatability assessment --- reproducibility, robustness, or replicability --- with all three being the strongest form of evidence (Level~4). &

Transparency itself does not establish causality (IV), measurement correctness (CV), or generalizability (EV). Independent verification that materials are accessible (Level~3) does not guarantee findings are correct. Reproducibility does not imply robustness or replicability --- these three dimensions are empirically partially independent and should not be assumed to predict one another \citep{nosek2026glossary,aczel2026}. A single successful replication does not guarantee universality. &

\textit{Level~1 --- Disclosure:} ``We report whether each open science practice was conducted, including AI model, version, parameters, and prompt protocol.'' \newline
\textit{Level~2 --- Sharing:} ``Data, code, and materials are deposited at [DOI]; pre-registration is at [URL].'' \newline
\textit{Level~3 --- Verification:} ``An independent audit confirmed all materials are accessible and code reproduces reported results.'' \newline
\textit{Level~4 --- Repeatability:} ``Results were [reproduced / robust to alternative specs / replicated with new data].'' \\

\end{longtable}
% ------------------------------------------------------------------
% Notes on relationships among validity principles
% ------------------------------------------------------------------
\vspace{1em}
\noindent\footnotesize\textbf{Notes on the relationships among validity principles.}
The five principles are parallel concerns, each addressing a distinct dimension of what can undermine the credibility of research conclusions. No single principle substitutes for the others, and gaps in different principles produce distinct types of problems. A causally identified effect on a poorly chosen outcome measure is a credible answer to the wrong question. Without adequate transparency and sharing, external parties cannot assess any of the other four principles. TRV modulates the strength of claims warranted by the other four principles: a perfectly transparent study with serious validity problems transparently produces evidence that is not credible. The three dimensions of repeatability at Level~4 (reproducibility, robustness, and replicability) are empirically partially independent \citep{nosek2026glossary,aczel2026} and should each be interpreted on their own terms. The appropriate strength of causal claims should scale with the overall validity profile.
\end{landscape}

% =====================================================================
% APPENDIX TABLE C: THREATS TO VALIDITY
% =====================================================================

\clearpage
\onecolumn
\begin{landscape}

\footnotesize
\setlength{\LTcapwidth}{14.5cm}
\setlength{\tabcolsep}{3pt}

\definecolor{principlebg}{gray}{0.80}
\definecolor{subbg}{gray}{0.92}

\begin{longtable}{%
  >{\RaggedRight\bfseries}p{2.0cm}%   Threat name
  >{\RaggedRight}p{5.3cm}%            Description
  >{\RaggedRight}p{7.2cm}%            Example in AI evaluation Research
}

\caption*{\textbf{Appendix C.} Threats to Validity in AI evaluation RCTs. Each validity principle is indicated by a shaded header row. Within the TRV section, sub-headers indicate which TRV level each threat belongs to. For each threat, the left column gives the threat name, the middle column gives a description, and the right column gives an illustrative AI evaluation RCT example. Abbreviations: CV = Construct Validity; IV = Internal Validity; EV = External Validity; SCV = Statistical Conclusion Validity; TRV = Transparency, Repeatability \& Verification.}
\label{tab:threats}\\

\toprule
\textbf{Threat} & \textbf{Description} & \textbf{Example in AI Evaluation Research} \\
\midrule
\endfirsthead

\multicolumn{3}{l}{\small\itshape Appendix C continued from previous page.}\\
\toprule
\textbf{Threat} & \textbf{Description} & \textbf{Example in AI Evaluation Research} \\
\midrule
\endhead

\midrule
\multicolumn{3}{r}{\small\itshape Continued on next page.}\\
\endfoot

\bottomrule
\endlastfoot

% ------------------------------------------------------------------
% PRINCIPLE HEADER: CV
% ------------------------------------------------------------------
\multicolumn{3}{>{\columncolor{principlebg}}p{\dimexpr2.0cm+5.3cm+7.2cm+4\tabcolsep\relax}}{\rule{0pt}{8pt}\textbf{Construct Validity (CV)} --- \textit{Whether outcome measures actually capture the theoretical constructs of interest, rather than proxies, artifacts, or constructs transformed by AI assistance itself.}}\\
\midrule

Inadequate construct explication &
Constructs are defined vaguely or incompletely, leaving ambiguity about what is being measured. &
A study claims to measure ``writing quality'' but operationalizes it only as word count or grammar score, without defining what quality means in an AI-assisted context. \\
\midrule

Construct confounding &
The operationalization captures more than one construct, making it impossible to attribute effects to the intended construct alone. &
Providing the treatment group with AI plus dedicated training means observed gains could reflect AI access, training, or their interaction --- the construct ``AI effect'' is confounded. \\
\midrule

Mono-operation bias (treatment) &
Only one version of the treatment is used, so findings may reflect the specific implementation rather than the broader construct. &
Testing only one AI model (e.g., GPT-4) and concluding about ``AI assistance'' in general, without considering that other models or prompting strategies may produce different effects. \\
\midrule

Mono-operation bias (outcome) &
Only one operationalization of the outcome is used, leaving construct breadth uncovered. &
Measuring programming ability with a single coding task; a one-item test cannot represent the full construct of programming competence. \\
\midrule

Mono-method bias &
All outcome measures rely on the same method (e.g., all self-report, or all automated metrics), conflating method variance with construct variance. &
Assessing ``code quality'' only through automated linting scores, missing dimensions like readability, maintainability, and correctness that require human judgment. \\
\midrule

Confounding constructs with levels &
Findings from a narrow range of construct levels are generalized to the full construct. &
Using simple, routine coding tasks (e.g., LeetCode ``easy'') and drawing conclusions about AI's impact on software engineering broadly. \\
\midrule

Reactive self-report changes &
Participants alter their responses due to social desirability, evaluation apprehension, or concern about how their answers will be perceived. &
Participants overreport satisfaction with AI tools because they want to appear tech-savvy, or underreport AI usage because they fear being seen as dependent on it. \\
\midrule

Experimenter expectancy &
Researchers' beliefs about expected outcomes bias how they interact with participants, evaluate outputs, or design study materials. &
A researcher enthusiastic about AI provides richer onboarding to the treatment group, or evaluates AI-assisted outputs more favorably when aware of condition assignment. \\
\midrule

Demand characteristics / hypothesis guessing &
Participants infer the study's hypothesis and adjust their behavior accordingly. &
Participants in the treatment group try harder because they know they have AI access and believe the study expects them to outperform controls. \\
\midrule

Treatment-sensitive factorial structure &
The factor structure of an outcome measure changes across conditions due to the intervention itself, meaning the construct being measured differs between groups. &
``Programming skill'' assessed before AI access involves syntax, logic, and debugging as distinct dimensions; with AI assistance, those dimensions may collapse into a single factor (task completion), making pre-post comparisons invalid. \\
\midrule

Multiple treatment interference &
Participants receive multiple overlapping treatments, making it impossible to isolate the effect of the focal intervention. &
Study participants also use Grammarly, web search, and other AI tools alongside the assigned tool, preventing attribution of effects to the specific AI system under study. \\

% ------------------------------------------------------------------
% PRINCIPLE HEADER: IV
% ------------------------------------------------------------------
\midrule
\multicolumn{3}{>{\columncolor{principlebg}}p{\dimexpr2.0cm+5.3cm+7.2cm+4\tabcolsep\relax}}{\rule{0pt}{8pt}\textbf{Internal Validity (IV)} --- \textit{Whether observed differences in outcomes can be attributed to the AI intervention rather than to alternative explanations such as confounding, selection, or concurrent events.}}\\
\midrule

Selection bias &
Treatment and control groups differ systematically at baseline on characteristics related to outcomes. &
Self-selection into AI use means early adopters tend to be more technically skilled or motivated than non-adopters, inflating apparent AI effects in observational studies. \\
\midrule

History / cointerventions &
Concurrent events or interventions --- other than the focal treatment --- affect outcomes during the study period. &
A major AI model update is released mid-study, changing control group behavior as participants begin using publicly available tools outside the study protocol. \\
\midrule

Maturation &
Participants improve over time due to practice, learning, or natural development independent of the AI intervention. &
A coding study runs over 8 weeks; participants in both groups improve simply from practice, making it difficult to isolate the AI effect from skill development. \\
\midrule

Attrition / differential dropout &
Participants drop out of the study, and dropout is related to condition or to outcomes, biasing the remaining sample. &
Participants in the control group who find the tasks most difficult are more likely to drop out, leaving a more skilled control group and inflating the apparent AI effect. \\
\midrule

Testing effects &
Repeated measurement sensitizes participants to the outcome construct or improves performance through practice with the measurement instrument itself. &
Participants complete the same coding assessment at baseline and follow-up; familiarity with the task format produces gains that are incorrectly attributed to AI access. \\
\midrule

Instrumentation &
The measurement instrument or procedure changes across timepoints or conditions, creating artificial differences. &
The AI interface is updated mid-study with a new user experience, changing how participants in the treatment group interact with the tool and confounding the treatment effect. \\
\midrule

Regression to the mean &
Participants selected for extreme baseline scores tend to score closer to the mean on subsequent measurements, regardless of intervention. &
A study recruits low-performing employees and finds that AI access ``improves'' their performance; part of the gain reflects natural regression rather than AI effect. \\
\midrule

Contamination &
Control group participants gain access to the treatment, reducing or eliminating the contrast between conditions. &
Control group participants learn about the AI tool from treated colleagues and begin using it informally, reducing or eliminating the intended treatment contrast. \\
\midrule

Interference / spillover &
Treatment of one participant affects the outcomes of others, violating the stable unit treatment value assumption (SUTVA). &
A treated employee shares AI-generated templates with untreated colleagues in the same team, improving control group performance and attenuating the apparent treatment effect. \\
\midrule

Resentful demoralization &
Control group participants, aware they are not receiving a potentially beneficial treatment, reduce their effort or motivation. &
Control group participants who know they lack AI access become discouraged and reduce effort, inflating the apparent AI treatment effect beyond its true value. \\
\midrule

Compensatory rivalry &
Control group participants, aware of the treatment, increase their effort to compensate, reducing the apparent treatment effect. &
Control participants told they are in the ``no AI'' condition work harder to prove they can match AI-assisted performance, compressing the estimated treatment effect. \\
\midrule

Model version drift &
The AI system changes during the study due to updates, fine-tuning, or parameter changes, introducing heterogeneity in the treatment itself. &
An API-accessed model is silently updated mid-study, changing output quality in ways not visible to the research team and creating within-condition heterogeneity. \\
\midrule

Differential compliance &
Participants in the treatment group do not use the AI as intended, or use it in ways that differ substantially from the protocol. &
Treatment group participants use the AI tool for only a fraction of tasks, or use it in ways not anticipated by the study design, weakening the intended treatment contrast. \\
\midrule

Non-equivalent control condition &
The control condition differs from the treatment condition in ways beyond AI access, confounding attribution. &
The control condition provides participants with no additional resources, while the treatment condition provides both AI access and a structured workflow, making it impossible to isolate the AI effect. \\
\midrule

Demand-supply mismatch &
The AI tool's capabilities do not match the demands of the assigned tasks, producing null effects that reflect task-tool mismatch rather than absence of AI impact. &
An AI coding assistant is tested on tasks requiring creative problem-solving that the model handles poorly; null results reflect tool-task misalignment rather than absence of AI impact. \\
\midrule

Novelty / Hawthorne effects &
Observed effects reflect participants' responses to novelty or being observed rather than to the AI intervention itself. &
Treatment group performance improves initially because participants are excited about using a new AI tool; effects decay as novelty wears off, but the study is too short to detect this. \\
\midrule

Confounded randomization unit &
The unit of randomization does not match the unit of analysis or the interference structure, producing biased estimates. &
Individuals within the same team are randomized independently, but team-level communication means interference is team-level; individual-level analyses underestimate standard errors and overstate precision. \\
\midrule

Ceiling / floor effects &
Outcome measures do not have sufficient range to detect effects at the extremes of the performance distribution. &
Expert participants are tested on tasks they already perform near-perfectly; AI access cannot improve scores that are already at the ceiling, producing null results that do not reflect absence of AI impact for non-expert populations. \\
\midrule

Insufficient follow-up &
The study does not measure outcomes at time points sufficient to capture delayed, sustained, or decay effects. &
A study measures outcomes immediately after a two-hour AI-assisted session; it cannot detect whether effects persist over weeks or whether skill transfer occurs after AI access ends. \\

% ------------------------------------------------------------------
% PRINCIPLE HEADER: EV
% ------------------------------------------------------------------
\midrule
\multicolumn{3}{>{\columncolor{principlebg}}p{\dimexpr2.0cm+5.3cm+7.2cm+4\tabcolsep\relax}}{\rule{0pt}{8pt}\textbf{External Validity (EV)} --- \textit{Whether findings generalize beyond the specific sample, tasks, settings, and AI system version studied.}}\\
\midrule

WEIRD sampling &
Samples drawn predominantly from Western, Educated, Industrialized, Rich, and Democratic populations limit generalizability to other demographic and cultural contexts. &
A study recruits participants exclusively from US university student pools; effects may not generalize to workers in non-WEIRD countries or to populations with different educational backgrounds and AI exposure. \\
\midrule

Convenience sampling &
Participants are recruited for accessibility rather than representativeness, limiting generalizability to target populations. &
An AI evaluation RCT study recruits participants through online platforms (e.g., Prolific, MTurk) who are atypically tech-savvy and experienced with online experiments compared to the target workforce population. \\
\midrule

Artificial task design &
Tasks used in the study differ substantially from real-world tasks in complexity, duration, stakes, or context. &
A study tests AI's effect on short, well-defined coding problems that bear little resemblance to the open-ended, collaborative, multi-day tasks that characterize professional software engineering. \\
\midrule

Ecological invalidity &
The study environment differs from the deployment context in ways that affect how participants interact with the AI. &
Participants complete tasks in a controlled experimental interface with no distractions, time pressure, or organizational context; real-world performance may differ substantially. \\
\midrule

Treatment specification mismatch &
The AI system studied differs from those used in deployment contexts in model family, version, configuration, or access modality. &
A study tests a specific, carefully configured AI system; findings may not transfer to off-the-shelf deployments where users interact with default configurations and less controlled prompting. \\
\midrule

Model version generalizability &
Findings from one model version may not generalize to subsequent versions due to capability changes, alignment updates, or behavioral shifts. &
Results from a study using GPT-4 may not apply to GPT-4o or later versions, even for similar tasks, because model behavior changes with updates. \\
\midrule

Short study duration &
Brief studies cannot capture long-term adaptation, skill transfer, dependency formation, or decay effects. &
A one-hour experiment shows AI impact for a coding task, but cannot determine whether workers develop sustainable new skills or become dependent on AI in ways that reduce capability when access is removed. \\
\midrule

Interaction of selection and treatment &
The effect of the AI intervention differs across subpopulations in ways that limit generalizability to populations with different characteristics. &
AI assistance strongly benefits novices but provides no benefit to experts; findings from a novice-heavy sample should not be generalized to expert practitioners without separate evidence. \\

% ------------------------------------------------------------------
% PRINCIPLE HEADER: SCV
% ------------------------------------------------------------------
\midrule
\multicolumn{3}{>{\columncolor{principlebg}}p{\dimexpr2.0cm+5.3cm+7.2cm+4\tabcolsep\relax}}{\rule{0pt}{8pt}\textbf{Statistical Conclusion Validity (SCV)} --- \textit{Whether the statistical evidence is sufficient and appropriately interpreted to support claimed effects.}}\\
\midrule

Low statistical power &
The study lacks sufficient sample size to reliably detect effects of plausible magnitude, producing high rates of false negatives and inflated effect size estimates when effects are detected. &
A study with 30 participants per condition has less than 20\% power to detect a moderate effect ($d{=}0.4$), meaning most true effects of that size will not be detected, and those that are detected will be substantially overestimated. \\
\midrule

Unreliable outcome measures &
Outcome measures have low reliability, attenuating observed effect sizes and reducing statistical power. &
A coding quality rubric with low inter-rater reliability (ICC $< 0.6$) introduces substantial measurement noise, attenuating effect size estimates and requiring larger samples to achieve adequate power. \\
\midrule

Violated statistical assumptions &
Analyses assume properties of the data (e.g., independence, normality, homoscedasticity) that do not hold, producing invalid inference. &
Participants within the same team share information and strategies, violating the independence assumption of standard regression; standard errors are underestimated and significance tests are anticonservative. \\
\midrule

Multiple comparisons without adjustment &
Testing many hypotheses without controlling the familywise or false discovery rate inflates the probability of spurious significant findings. &
A study tests AI causal effect across 12 outcome variables and 4 subgroups without multiplicity correction; at least one spuriously significant finding is expected by chance even under the null. \\
\midrule

Researcher degrees of freedom &
Undisclosed analytical flexibility --- in outcome selection, covariate inclusion, exclusion criteria, or analysis timing --- inflates false positive rates. &
A researcher analyzes data multiple ways and reports only the specification that yields a significant result, without disclosing the alternatives explored. \\
\midrule

Lack of established benchmarks &
The field lacks consensus on what constitutes practically meaningful effect sizes for AI impact, making it difficult to interpret the practical significance of findings. &
A study reports $d{=}0.3$ without any basis for determining whether this effect is large enough to justify implementation costs, because no domain-specific benchmarks exist. \\
\midrule

HARKing &
Hypotheses are formulated or modified after examining data and presented as if they were pre-specified, inflating apparent confirmatory evidence. &
Exploratory subgroup analyses conducted post-hoc are reported as pre-specified confirmatory analyses, overstating the evidential value of the findings. \\
\midrule

Selective outcome reporting &
Only statistically significant or favorable outcomes are reported, biasing the published literature toward positive findings. &
A study collects 8 outcome measures but reports only the 2 that showed significant treatment effects, without disclosing that others were collected. \\
\midrule

Underpowered interaction tests &
Subgroup and moderation analyses require substantially larger samples than main effect tests, but this is rarely reflected in sample size planning. &
A study powered to detect a main effect of $d{=}0.4$ tests whether the effect differs between novices and experts but has only 10\% power to detect a plausible interaction, generating uninformative null moderation results. \\

% ------------------------------------------------------------------
% PRINCIPLE HEADER: TRV
% ------------------------------------------------------------------
\midrule
\multicolumn{3}{>{\columncolor{principlebg}}p{\dimexpr2.0cm+5.3cm+7.2cm+4\tabcolsep\relax}}{\rule{0pt}{8pt}\textbf{Transparency, Repeatability \& Verification (TRV)} --- \textit{Whether the research process and findings can be understood, independently verified, and shown to hold when aspects of the research process are repeated.}}\\
\midrule

% --- Level 1 sub-header ---
\multicolumn{3}{>{\columncolor{subbg}}p{\dimexpr2.0cm+5.3cm+7.2cm+4\tabcolsep\relax}}{\rule{0pt}{6pt}\textit{Disclosure failures (Level~1)}}\\
\midrule

Selective reporting &
Only significant or favorable results are published or disclosed, creating a distorted picture of the evidence base. &
A research team runs three AI evaluation RCT studies; only the one showing positive results is published, while null results remain unpublished. \\
\midrule

Incomplete methods description &
Insufficient detail to allow others to evaluate or reproduce the study's procedures. &
The study does not report the specific model name and version, the exact prompts used, parameter settings, or the human-AI interface design, making replication impossible. \\
\midrule

Undisclosed researcher degrees of freedom &
Analytical decisions made during data processing and analysis are not reported, preventing assessment of how sensitive findings are to those choices. &
The paper does not disclose that several alternative outcome definitions were considered, or that the primary analysis was changed after an interim look at the data. \\
\midrule

Conflict of interest non-disclosure &
Financial or other interests that could bias design, analysis, or reporting are not disclosed. &
Researchers whose employment or funding depends on demonstrating AI effectiveness do not disclose this relationship, preventing readers from assessing potential for motivated reporting. \\
\midrule

Undisclosed prompt engineering &
The prompts used to elicit AI responses are not reported, even though small variations can substantially affect outputs and outcomes. &
A study reports that participants used ``an AI writing assistant'' without disclosing the specific instructions given to the AI or how the human-AI interaction was structured. \\
\midrule

% --- Level 2 sub-header ---
\multicolumn{3}{>{\columncolor{subbg}}p{\dimexpr2.0cm+5.3cm+7.2cm+4\tabcolsep\relax}}{\rule{0pt}{6pt}\textit{Sharing failures (Level~2)}}\\
\midrule

Inaccessible materials &
Data, code, or study materials are nominally reported as available but cannot actually be accessed, used, or verified by others. &
Raw participant data and analysis code are not deposited in a public repository, or links are broken, or the repository requires permissions that are not granted, making it impossible for others to verify reported results or conduct re-analyses. \\
\midrule

AI system opacity &
Key properties of the AI system used are unknown, undisclosed, or inaccessible to other researchers. &
A study uses a proprietary model accessed via API without reporting model version, temperature settings, system prompt, or the organization's internal fine-tuning, preventing meaningful replication. \\
\midrule

Missing version control &
Analytical code, study materials, and interfaces are not version-controlled, making it impossible to identify what was used at any given point in the study. &
Analysis code is modified multiple times after data collection begins, with no record of which version was used for which analysis, creating unreproducible results. \\
\midrule

Data privacy and proprietary constraints &
Privacy requirements or organizational confidentiality agreements prevent sharing data or materials that would otherwise be necessary for verification. &
A study conducted with proprietary organizational data cannot share participant-level data, limiting what the community can verify and narrowing the appropriate strength of claims. \\
\midrule

% --- Level 3 sub-header ---
\multicolumn{3}{>{\columncolor{subbg}}p{\dimexpr2.0cm+5.3cm+7.2cm+4\tabcolsep\relax}}{\rule{0pt}{6pt}\textit{Verification failures (Level~3)}}\\
\midrule

Deviation from pre-registered protocol &
The study deviates from the pre-registered design without documentation or justification, undermining the pre-registration's purpose. &
The primary outcome measure is changed after data collection begins due to ceiling effects, but the deviation is not disclosed in the published report. \\
\midrule

Undocumented human-AI interaction &
Actual usage patterns differ from the described protocol, and deviations are not documented. &
Participants develop idiosyncratic interaction strategies not captured in the protocol; these strategies contribute to outcome variation but are invisible in the published report. \\
\midrule

System evolution &
The AI system changes over time, making previously reported results difficult or impossible to reproduce. &
The API endpoint used in the study no longer returns the same outputs because the underlying model has been updated, and the original model version is no longer accessible. \\
\midrule

Insufficient transparency to calibrate claims &
The overall level of transparency is insufficient for the strength of claims being made; strong claims require more verifiability than weak ones. &
A study claiming that AI ``definitively improves cognitive performance'' provides only a brief methods section with no shared data, code, or pre-registration, making the strong claim unverifiable. \\
\midrule

% --- Level 4 sub-header ---
\multicolumn{3}{>{\columncolor{subbg}}p{\dimexpr2.0cm+5.3cm+7.2cm+4\tabcolsep\relax}}{\rule{0pt}{6pt}\textit{Repeatability failures (Level~4)}}\\
\midrule

Non-reproducibility &
Independent parties cannot confirm the reported results by rerunning the analysis on the shared data and code --- either because materials are incomplete, code does not run, or results differ from those reported. &
A researcher shares data and code but the analysis script contains errors or undocumented dependencies; independent analysts obtain different numerical results from the same inputs. \\
\midrule

Analytical fragility (non-robustness) &
Findings depend critically on specific analytical choices such that alternative justifiable analyses yield substantially different or contradictory conclusions. &
A study reports $d{=}0.4$, but independent reanalysis finds that excluding outliers or adjusting for baseline differently reduces the effect to $d{=}0.1$ or reverses its direction. Large-scale evidence suggests only about one-third of reanalyses closely match original quantitative estimates \citep{aczel2026}. \\
\midrule

Non-replicability &
Independent teams testing the same research question with new data obtain substantially different results, suggesting the original finding does not generalize beyond its specific study context. &
A study showing strong AI effects in a controlled lab setting with student participants fails to replicate when an independent team tests the same question with professional workers in naturalistic settings. Evidence from the social and behavioral sciences suggests replication effect sizes are on average less than half the magnitude of original estimates \citep{tyner2026}. \\

\end{longtable}

\end{landscape}

% =====================================================================
% APPENDIX D: GLOSSARY OF KEY TERMS
% =====================================================================

\clearpage
\onecolumn
\begin{landscape}

\footnotesize
\setlength{\LTcapwidth}{21cm}
\setlength{\tabcolsep}{4pt}

\begin{longtable}{%
  >{\RaggedRight}p{3.6cm}%   Term
  >{\RaggedRight}p{7.4cm}%   Definition as used in this paper
  >{\RaggedRight}p{9.2cm}%   Notes on variant usage and potential confusion
}

\caption*{\textbf{Appendix D.} Glossary of Key Terms. This table defines terms as used in this paper and notes common variant usages that create confusion in the literature. Terms are grouped thematically rather than alphabetically to reflect conceptual relationships.}
\label{tab:glossary}\\

\toprule
\textbf{Term} & \textbf{Definition as used in this paper} & \textbf{Notes on variant usage and potential confusion} \\
\midrule
\endfirsthead

\multicolumn{3}{l}{\small\itshape Appendix D continued from previous page.}\\
\toprule
\textbf{Term} & \textbf{Definition as used in this paper} & \textbf{Notes on variant usage and potential confusion} \\
\midrule
\endhead

\midrule
\multicolumn{3}{r}{\small\itshape Continued on next page.}\\
\endfoot

\bottomrule
\endlastfoot

% ------------------------------------------------------------------
% BLOCK 1: The "evaluation" / "evals" family
% ------------------------------------------------------------------
\rowcolor{gray!12}
\multicolumn{3}{l}{\textbf{AI evaluation: a contested umbrella and its subtypes}} \\
\midrule

\textbf{Evaluation (AI context)}
&
An umbrella term for any systematic assessment of an AI system's capabilities, behaviors, safety properties, or real-world effects. AI evaluation encompasses many distinct methods with different designs, inferential capacities, and objects of assessment. This paper focuses on one specific subtype: randomized controlled trials measuring effects on human outcomes.
&
``Evaluation'' is frequently used as though it refers to a single method or a single inferential standard. In practice it collapses a wide range of study types that differ substantially in rigor, design, and what they can conclude. The term ``AI evaluation'' is sometimes used interchangeably with ``benchmark evaluation,'' which is a much narrower usage. See the rows below for distinctions among the main subtypes.
\\

\midrule

\textbf{Evals (industry usage)}
&
A colloquial shorthand used by AI developers and safety researchers for structured assessments of model behavior, most commonly referring to automated benchmark-style tests or structured adversarial exercises. In industry usage, ``evals'' most frequently means benchmark evaluations.
&
The term ``evals'' is used differently across organizations. In many AI safety discussions it implies benchmark evaluations specifically. In others it serves as a broad umbrella covering red-teaming, human studies, and structured capability assessments. This ambiguity matters: treating ``evals'' as synonymous with benchmarks implicitly excludes the human-centered, causal study designs this paper addresses. A model that ``passes evals'' has not necessarily been evaluated in ways that support causal claims about human outcomes.
\\

\midrule

\textbf{Benchmark evaluation}
&
An assessment of model outputs against a fixed test set, typically automated and without human participants in the role of subjects. Benchmarks measure model performance under standardized conditions and are designed to be repeatable across models and over time.
&
Benchmarks do not randomize human participants and therefore do not support causal inference about effects on human behavior or performance. A model scoring highly on a benchmark does not imply that users of that model will perform better on real tasks. Benchmark scores are also susceptible to dataset contamination (training data leakage) and Goodhart's Law: once a benchmark becomes a target, it becomes a less valid measure of the underlying capability \citep{weidinger2025}.
\\

\midrule

\textbf{Red-teaming}
&
A structured adversarial exercise in which human testers (or automated agents) attempt to elicit harmful, unsafe, or policy-violating behavior from an AI system. Red-teaming focuses on identifying failure modes under adversarial conditions rather than characterizing average-case performance.
&
Red-teaming is dynamic and adversarial, in contrast to benchmarks which use fixed inputs. Red-teaming exercises do not randomize human participants as subjects and do not produce estimates of population-level causal effects on human performance. Findings characterize worst-case or tail-risk behavior, not the distribution of outcomes that a typical user would experience. The two methods are complementary, not interchangeable \citep{weidinger2025}.
\\

\midrule

\textbf{AI evaluation RCT}
&
A randomized controlled trial in which human participants are randomly assigned to conditions with and without AI access, and outcomes are measured to estimate the causal effect of that access on human behavior, performance, or wellbeing. This is the primary study type addressed by our framework.
&
AI evaluation RCTs are distinct from benchmarks (which assess model outputs without randomizing human participants), from red-teaming (which is adversarial rather than population-representative), and from observational studies (which cannot support causal claims). The inferential bar is higher, but so is the relevance to deployment and governance decisions about AI effects on people.
\\

\midrule

% ------------------------------------------------------------------
% BLOCK 2: The "uplift" family
% ------------------------------------------------------------------
\rowcolor{gray!12}
\multicolumn{3}{l}{\textbf{The ``uplift'' family: a contested and polysemous term}} \\
\midrule

\textbf{Uplift (AI evaluation contexts, general)}
&
In AI evaluation, ``uplift'' most commonly refers to a change in human capability or performance attributable to AI access. This usage appears across both AI safety and human performance research, though with different scope and valence in each (see rows below). We avoid the term as a primary label in this paper because it carries a directional implication (improvement) that the evidence does not always support, and because it conflates distinct research questions across communities.
&
The core problem with ``uplift'' as a label is that it presupposes a direction of effect. AI access may improve performance, leave it unchanged, or reduce it, depending on the task, the population, the system, and the context. Using ``uplift'' implicitly treats improvement as the default or expected outcome, which biases how studies are designed, analyzed, and reported. A treatment effect, by contrast, is neutral: it describes whatever change occurred, in whatever direction.
\\

\midrule

\textbf{Uplift (AI safety context)}
&
In AI safety research, ``uplift'' refers to an increase in human capability resulting from AI access, with particular focus on whether that capability increase creates or amplifies risk. Discussions of CBRN (chemical, biological, radiological, nuclear) threats often center on whether AI provides uplift to actors seeking to cause harm, but the concept also applies to beneficial capability gains. The directional implication (AI makes someone more capable) is consistent across safety applications; whether that is good or bad depends on who benefits and to what end \citep{aisi2024}.
&
A common shorthand is to frame uplift in AI safety as being specifically about bad actors or misuse. This is partially accurate for CBRN and security applications, but the underlying concept (does AI access increase human capability in domain X) is the same question asked in human performance research. The communities use the same word for the same underlying phenomenon but in different risk frames. Conflating the safety usage with the human performance usage leads to confusion about whether ``uplift studies'' are measuring beneficial effects, harmful capability gains, or both.
\\

\midrule

\textbf{Uplift (human performance context)}
&
In human performance research and documents such as the RAND GPAI evaluation framework, ``uplift'' refers to measurable increases in human performance or output resulting from AI access, without implying any particular use case or risk frame \citep{rand2025}.
&
This usage shares the directional assumption of the safety usage (uplift means capability went up) but applies it in a productivity or task-completion context rather than a risk context. Like the safety usage, it does not require randomization and is sometimes applied to pre/post or observational designs, which means ``uplift'' in this sense says nothing about whether causal identification requirements were met.
\\

\midrule

\textbf{Human uplift study}
&
A colloquial label for studies assessing whether AI assistance changes human performance. We avoid this as a primary term because it specifies neither the study design, the estimand, nor the direction of effect.
&
Some researchers and AI organizations use ``human uplift study'' and ``AI evaluation RCT'' interchangeably, but the two are not the same. An AI evaluation RCT has a specific design (randomization of human participants), a specific inferential goal (causal estimation of treatment effects), and no built-in directional assumption about the outcome. We recommend replacing ``human uplift study'' with design-specific language: ``an RCT estimating the effect of [AI system] on [human outcome] for [population] compared with [counterfactual].''
\\

\midrule

\textbf{Uplift modeling (causal ML / industry)}
&
A machine learning approach for estimating heterogeneous treatment effects at the individual or subgroup level, typically applied to data from randomized experiments. The goal is to identify which individuals are most likely to respond to a treatment and optimize targeting accordingly \citep{gutierrez2017}.
&
Uplift modeling is an analytical technique applied after collecting experimental or observational data, to model how effects vary across individuals; it presupposes an experiment rather than replacing one. Where a standard A/B test yields an average treatment effect, uplift modeling estimates conditional average treatment effects (CATE) to support targeting decisions. The word ``uplift'' in this context refers to individual response to an intervention and is unrelated to AI safety or human performance AI evaluation usage.
\\

\midrule

% ------------------------------------------------------------------
% BLOCK 3: Causal inference and estimands
% ------------------------------------------------------------------
\rowcolor{gray!12}
\multicolumn{3}{l}{\textbf{Causal inference and estimands}} \\
\midrule

\textbf{Treatment effect}
&
The difference in an outcome for a unit under treatment versus the same unit under the counterfactual of no treatment. Because both states cannot be observed simultaneously, treatment effects are estimated using comparison groups. A treatment effect can be positive, negative, or null.
&
Unlike ``uplift,'' ``treatment effect'' carries no directional assumption. This paper uses it as the preferred term for the quantity estimated in an AI evaluation RCT. See \hyperref[tab:inferential]{Appendix~B} for the inferential scope of each validity principle.
\\

\midrule

\textbf{Average treatment effect (ATE)}
&
The expected difference in outcomes if all units in a population were assigned to treatment versus if all were assigned to control. The ATE is the estimand most commonly targeted by RCTs with full compliance.
&
The ATE describes a population-level average that may mask substantial heterogeneity. In AI evaluation, average effects can be near-zero while some subgroups experience large positive or negative effects. Our framework emphasizes heterogeneity analysis alongside primary ATE estimates for this reason.
\\

\midrule

\textbf{Intent-to-treat (ITT) effect}
&
The average effect of being \textit{assigned} to a treatment condition, regardless of whether participants actually used the treatment. In AI evaluation RCTs, ITT captures the effect of offering AI access, including the effects of non-adoption and partial use.
&
ITT and the complier average causal effect (CACE) answer different questions and should be reported separately. ITT is appropriate for policy questions about deploying AI at scale where uptake will be variable; CACE addresses effects among those who actually engage with the system \citep{shadish2002}.
\\

\midrule

\textbf{Complier average causal effect (CACE) / Local average treatment effect (LATE)}
&
The average treatment effect among participants who comply with their assigned condition. In AI evaluation RCTs, this means those who actually used the AI system. CACE is estimated using randomized assignment as an instrument for actual use.
&
CACE is labeled LATE in the econometrics literature; both terms refer to the same estimand under an instrumental variables framework. Reporting both CACE and ITT provides a fuller picture of AI system effects than either alone.
\\

\midrule

\textbf{Estimand}
&
The specific quantity a study is designed to estimate, defined before data collection. A complete estimand specification identifies the population, treatment, comparator, outcome, and any conditioning variables.
&
Without a pre-specified estimand, researchers retain post-hoc flexibility to shift between ITT, CACE, ATE, and subgroup effects after observing data, inflating false-positive risk. This is one of the most consequential researcher degrees of freedom in AI evaluation studies. See \hyperref[tab:guidelines]{Appendix~A}, Guideline~1.
\\

\midrule

\textbf{Counterfactual / comparator}
&
The condition against which AI-assisted performance is compared. A well-specified AI evaluation RCT defines the comparator explicitly: AI-unassisted performance, performance with a legacy tool, or performance with a different AI system.
&
``Baseline'' is used loosely to mean either a pre-treatment measurement of the same participants or a control group without AI access. These are not equivalent. A pre-treatment measurement does not substitute for a randomized control group and does not support causal inference about AI effects.
\\

\midrule

% ------------------------------------------------------------------
% BLOCK 4: Claim types
% ------------------------------------------------------------------
\rowcolor{gray!12}
\multicolumn{3}{l}{\textbf{Claim types: what a given study design can support}} \\
\midrule

\textbf{Descriptive claim}
&
A statement about what occurred in a specific sample under specific conditions, without implying causation or generalization. Example: ``Participants in the AI-assisted group produced longer responses on average.''
&
Benchmark evaluations and observational studies typically support descriptive claims at most. Framing descriptive findings causally or as generalizable, without randomization and appropriate external validity, is a pervasive error in AI evaluation.
\\

\midrule

\textbf{Correlational claim}
&
A statement about statistical association between AI use and an outcome, without implying that AI access caused the outcome. Usage logs, deployment data, and cross-sectional surveys support correlational claims.
&
Correlational findings can motivate RCT designs but do not establish causation. Omitted variable bias, reverse causation, and selection effects are the primary threats to causal interpretation of correlational AI data.
\\

\midrule

\textbf{Causal claim}
&
A statement that AI access or use caused a change in a specified outcome for a specified population, supported by a design that enables causal inference. In this paper, causal claims require randomization and attention to all five validity principles.
&
Causal claims vary in scope. An internally valid RCT supports a causal claim for the study sample under study conditions; external validity determines how far that claim generalizes. See \hyperref[tab:inferential]{Appendix~B} for inferential scope by validity principle.
\\

\end{longtable}

\end{landscape}
\end{document}